\documentclass[useAMS,usenatbib]{mnras}
\usepackage{amsmath}
\usepackage{float}
\usepackage{graphicx}
\usepackage{txfonts}
\usepackage{indentfirst}
\usepackage{color}
\usepackage{ulem}
\usepackage{hyperref}
\usepackage[Symbolsmallscale]{upgreek}
\usepackage{nomencl}
\usepackage{tabularx}
\usepackage[table,xcdraw]{xcolor}
\usepackage{subcaption}
\usepackage{mathbbol}

\newcommand{\eq}[1]{\begin{equation}  #1 \end{equation}}
\newcommand{\eqs}[1]{\begin{equation} \begin{split} #1 \end{split} \end{equation}}

\newcommand{\br}[1]{\left( #1 \right)}

\newcommand{\bb}[1]{\left[ #1 \right]}

\newcommand{\dd}{{\rm d}}

\newcommand{\vek}[1]{\mbox{\boldmath $#1$}}

\newcommand{\svek}[1]{\mbox{\boldmath \scriptsize $#1$}}

\newcommand{\ic}{{\rm i}}

\def\araa{ARA\&A}
\def\apj{ApJ}
\def\apjl{ApJ}
\def\apjs{ApJS}
\def\aap{A\&A}

\def\mnras{MNRAS}
\def\nat{Nature}

\title[Plasma Microlensing]{Plasma microlensing dynamic spectrum probing fine structures in the ionized interstellar medium}

\author[Shi, Xu]{Xun Shi$^{1}$\thanks{E-mail: xun@ynu.edu.cn}, Zhu Xu$^{1}$\thanks{E-mail: zhu@mail.ynu.edu.cn} \\
$^{1}$South-Western Institute for Astronomy Research (SWIFAR), Yunnan University, 650500 Kunming, P. R. China}

\begin{document}
\maketitle
\label{firstpage}
\begin{abstract}
Gravitational microlensing has become a mature technique for discovering small gravitational lenses in the Universe which are otherwise beyond our detection limits. 
Similarly, plasma microlensing can help us explore cosmic plasma lenses. 
Both pulsar scintillation and extreme scattering events of compact radio sources suggest the existence of
$\sim$ AU scale plasma lenses in the ionized interstellar medium (IISM), whose astrophysical correspondence remains a mystery.
We demonstrate that plasma microlensing events by these plasma lenses recorded in the form of wide-band dynamic spectra are a powerful probe of their nature.
Using the recently developed Picard-Lefschetz integrator for the Kirchhoff-Fresnel integral, we simulate such dynamic spectra for a well-motivated family of single-variable plasma lenses.
We demonstrate that the size, strength, and shape of the plasma lens can be measured from the location of the cusp point and the shape of spectral caustics respectively,
with a combination of distances and the effective velocity known a priori or measured from the widths of the interference pattern.
Future wide-band observations of pulsars, whose plasma microlensing events may be predictable from parabolic arc monitoring, are the most promising ground to apply our results
for a deeper insight into the fine structures in the IISM. 
\end{abstract}


\section{Introduction}

Plasma inhomogeneities form lenses. 
Radio wave traveling in the ionized interstellar medium (IISM) gets diffracted and refracted by these plasma lenses. 
As a result, compact radio sources scintillate, i.e. their flux varies with time. 
The best regular probe of IISM scintillation is the pulsar dynamic spectrum, recording of pulsar flux variations over time and frequency \citep[e.g.][]{rickett69, rickett84, cordes86, gupta94, johnson98}. 
In addition, `extreme scattering events' (ESEs) are observed in some compact radio sources as rare, large variations in flux density over days to weeks are also a manifestation of IISM scintillation \citep{fiedler87, fiedler94, lazio01, bannister16}. They provide unique constraints to the properties of the small-scale IISM structures.

Traditionally, IISM density fluctuations as a result of the ubiquitous turbulent motions were thought to be responsible for most of the scintillation (\citealt{rickett90, narayan92}, and references therein) despite the hint of coherent plasma lenses by the few observations of ESEs \citep{romani86}. 
However, since the turn of the century, new observations have triggered a paradigm shift. 
Firstly, the discovery of a distinct parabolic arc in pulsar secondary spectra i.e. square amplitude of the Fourier-transformed dynamic spectra \citep{stinebring01}, and sometimes multiple parabolic arcs \citep{putney06}, have suggested the existence of discrete scattering screens between the pulsar and the observer, and pointed to the anisotropic nature of the scattering \citep{walker04, cordes06}. 
Secondly, the detection of inverted arclets on top of the main parabolic arc reveals discrete structures on the scattering screen which are found to be long-lived from long-term monitoring \citep{hill03, hill05}.
With the accumulation of high-quality pulsar dynamic spectra, parabolic arcs and arclets turn out to be very common. 
This leads to the idea that the discrete structures in the IISM could be the major contributor to IISM scintillation. 
They act as plasma lenses, which could be the same as those responsible for the ESEs.

What are these plasma lenses remains a great mystery. 
One difficulty of resolving this mystery is the small estimated sizes of the plasma lenses, which range from sub-AU as inferred from pulsar secondary spectra \citep{pen14b}, to AU and tens of AU from ESE observations. 
Direct observational evidence of these fine IISM structures is thus hard to find. 
Theoretically, several ideas have been proposed, including corrugated reconnection sheet \citep{pen12, pen14b, liu16, simard18}, ionized skins of molecular clumps \citep{walker17}, and filaments ionized by hot stars \citep{walker17}. 
Most of them invoke a grazing viewing angle to mitigate the need for extremely high density fluctuations suggested by early studies assuming spherical symmetry \citep{romani87, clegg98}.

Reconstruction of the scattered images with pulsar secondary spectra provides essential information on the geometry of the lens. 
In a detailed VLBI study of a bright pulsar PSR B0834+06 \citep{brisken10}, much of the power on the main parabolic arc is resolved into many discrete inverted arclets, which correspond to scattered images lining up along a thin line that crosses the origin. 
The high degree of anisotropy implies that the lens plane deflects light mainly in one out of the two dimensions on the plane of the sky. 
Another somewhat under-appreciated piece of evidence is that the arclets stay on the main parabolic arc whose apex remains at the origin as the relative locations of the source, lens, and observer change with time, 
which suggests that this line of images always crosses the source line of sight \citep{shi21b}. 
This implies that the lens plane is unlikely composed of small discrete lenses lined up in a filamentary geometry, 
but more likely a two-dimensional wave that is mainly fluctuating in one direction while flat in the perpendicular direction. 
This preferred geometry is in accord with that proposed by \citet{pen14b} where the arclets correspond to images created at the folds of a wavy sheet viewed at a grazing angle.

Microlensing events, i.e. flux variations of a source when a lens passes through its line of sight, offer a complimentary test of the properties of the plasma lens. 
ESEs are usually microlensing events where the flux of compact radio sources is recorded in one or several frequency bands.   
Traditionally, ESEs are detected in observations of active galactic nuclei. 
ESEs from these extragalactic sources have a downside of being very rare. 
For example, only $\sim$15 events have been discovered in a search of ESEs with 1200 source-years of archival observations \citep{fiedler94}.
Pulsars are potentially better sources for plasma microlensing events given their small angular sizes and fast proper motions.
Pulsar ESEs have recently been detected as long term variations in the dispersion measure, pulsar flux, scintillation strength, and scintillation bandwidth \citep{maitia03, coles15, kerr18},
from which the sizes of plasma lenses are inferred to be also around $\sim$AU. 

If the arclets in pulsar secondary spectra are indeed created by discrete plasma lenses, these lenses should also cause plasma microlensing events when they traverse the pulsar line of sight.
These events should occur at a much higher rate given the high number of arclets observed in some systems, 
and may have a much shorter duration given the small estimated lens sizes. 
Moreover, the occurrence time of such events can be predicted by monitoring the motion of arclets along a main parabolic arc, 
enabling real-time wide-band and VLBI observations which are greatly valuable for studying the underlying lens structure.
In particular, ultra-wide-band observations of pulsar plasma microlensing events, which have been feasible only very recently thanks to the new 0.7-4 GHz receiver on the Parkes telescope \citep{johnston21}, 
could revolutionize the study of the IISM. 

Not in pace with the observational developments, there is so far a lack of theoretical predictions of wide-band signatures of plasma microlensing events. 
This motivates this work where we simulate the wide-band dynamic spectra of plasma microlensing events and study how the properties of the lens can be inferred from them.
We introduce plasma lensing theory in Section.\;\ref{sec:lensing}, present the method and results of dynamic spectra simulation in Section.\;\ref{sec:sim}, demonstrate how to contrain the lens properties in Section.\;\ref{sec:constrain}, and conclude in Section.\;\ref{sec:conclusion}.

\section{Plasma lenses and plasma lensing}
\label{sec:lensing}
\subsection{Plasma Inhomogeneity as Lenses}
When the light from a source at distance $D$ travels across some deflecting medium of thickness $\ell_{\rm d}$, it gets a phase shift due to the fluctuating electron number density $n_{\rm e}$ in the deflecting medium 
\eq{
\Delta \Phi_{\rm disp}  =  2\uppi (n - 1) \nu \ell_{\rm d} / \it{c}  \,,
}
where $\nu$ is the frequency of the signal and $\it{c}$ the speed of light. The refractive index $n$ is related to the electron number density via the plasma frequency $\nu_{\rm p}$ as
\eqs{
\label{eq:ne2n}
n - 1 & \approx - \frac{1}{2} \frac{\nu_{\rm p}^2}{\nu^2} \\
& =  - 4.05 \times10^{-11} \frac{n_{\rm e}}{1\, \rm cm^{-3}} \br{\frac{\nu}{\rm 1\, GHz}}^{-2} \,.
}
The intervening plasma leads to a negative $\Delta \Phi_{\rm disp}$ due to the superluminal phase velocity in a plasma and a corresponding refractive index $n<1$.

This phase delay leads to distortion of the wavefront, and subsequently, a deviation of the light travel path from a straight line. 
This further introduces a geometrical phase delay $\Delta\Phi_{\rm geo}$. 
When the deflection material occupies a small fraction of the line-of-sight distance, it can be considered to lie on a thin scattering screen. 
In this case, and under small-angle approximation that is valid to high precision for IISM scintillation studies,
\eq{
\Delta\Phi_{\rm geo} = \tau_{\rm norm} \nu \frac{|\vek{\theta} - \vek{\beta}|^2}{2}
}
with $\vek{\theta}$ being the angular coordinates on the scattering screen i.e. the lens plane, $\vek{\beta}$ being the angular coordinates on the source plane.
\eqs{
\label{eq:tau_norm}
\tau_{\rm norm} & = \frac{2\uppi}{\it{c}} \frac{1 - f_{\rm d}}{f_{\rm d}} D \\
& = 1.52 \times 10^4 \frac{1 - f_{\rm d}}{f_{\rm d}} \frac{D}{\rm{kpc}} \,\bb{{\rm mas^{-2}\; GHz^{-1}}}
} 
is an effective distance depending on the distance $D$ from the pulsar to observer, and the fractional distance $f_{\rm d}$ between the pulsar and the deflecting medium with respect to $D$. Physically, the prefactor $\tau_{\rm norm} \nu$ defines a characteristic angular scale on the lens plane: the Fresnel scale $\theta_{\rm F} = 1 / \sqrt{\tau_{\rm norm} \nu}$. For typical configuration of a Milky Way pulsar, $\theta_{\rm F}$ is only a fraction of a milli-arcsecond (mas).

The total phase delay 
\eq{
\label{eq:phase_delay}
\Delta\Phi = \Delta\Phi_{\rm geo} + \Delta \Phi_{\rm disp} = \tau_{\rm norm} \nu  \bb{\frac{|\vek{\theta} - \vek{\beta}|^2}{2} - \psi}
}
is the key quantity governing the deflected light propagation from the source to the observer. 
Here, we have defined the deflection potential $\psi \equiv - \Delta \Phi_{\rm disp} / (\tau_{\rm norm} \nu)$ following the gravitational lensing convention.
In fact, the Fermat potential in the gravitational lensing literature \citep{schneider85},
\eq{
\tau(\vek{\theta}, \vek{\beta}) = \frac{|\vek{\theta} - \vek{\beta}|^2}{2} - \psi(\vek{\theta})
}
is just a convenient dimensionless form of $\Delta\Phi$, with the first and second term corresponding to the geometric phase delay and the phase delay occurred directly at the lens, respectively \footnote{In gravitational lensing, Fermat potential is a measure for both phase delay and time delay; In the dispersive plasma lensing, the Fermat potential reflects phase delay but not time delay, as the latter depends on group velocity and the former on phase velocity. }. 
In our case where the lens is the plasma, the deflection potential $\psi$ can be expressed as
\eq{
\label{eq:psi_phys}
\psi  = 8.35 \times10^{-3}  \br{\frac{N_{\rm e}}{\rm cm^{-3} \times AU}} \br{\frac{\nu}{\rm GHz}}^{-2} \br{\frac{f_{\rm d}}{1 - f_{\rm d}}} \br{\frac{D}{\rm kpc}}^{-1}  \, \bb{\rm mas^2}
}
where $N_{\rm e} = n_{\rm e} \ell_{\rm d}$ is the column density of the plasma lens. The positive sign of $\psi$ and thus negative sign of $\Delta \Phi_{\rm disp}$ implies that a lens composed of plasma overdensity diverges light rather than converging it, as opposed to a gravitational lens.

Consider the lens to be at a fixed location, the angular coordinate of the source is sufficient to specify the relative orientations of the source, lens, and the observer. 
Thus, we can directly use $\vek{\beta}$ as the time coordinate of the microlensing event. 
The two are related as $t = - \beta  D (1 - f_{\rm d})/ V_{\rm sr}$, 
where $\vek{V}_{\rm s,r}$ is the relative velocity of the scattering screen with respect to the pulsar-observer line of sight, and $\beta$ is the component of $\vek{\beta}$ in the direction of $\vek{V}_{\rm s,r}$.
The wave field on the observer's plane as a function of $\vek{\beta}$ is given by the Kirchhoff-Fresnel integral \citep[e.g.][]{walker04}
\eq{
\label{eq:KF}
\vek{E}(\vek{\beta}, \nu) = \frac{1}{2\uppi \ic \theta_{\rm F}^2} \int \dd^2 \vek{\theta} \exp{\bb{\ic \Delta\Phi(\vek{\beta}, \vek{\theta}, \nu)}}
}
up to a normalization constant. The primary observable of plasma lensing is the dynamic spectrum i.e. source intensity $I = \vek{E} \vek{E}^*$ as a function of time and frequency.

\subsection{Plasma Lensing Framework}
In general, the Kirchhoff-Fresnel integral (Eq.\;\ref{eq:KF}) has to be performed to compute the dynamic spectrum given a lens configuration. For some purposes, it is adequate to approximate the integral by a sum over a finite number of decrete points where the phases are stationary \citep[e.g.][]{walker04}
\eq{
\label{eq:stationary}    
 \nabla \Phi \propto \nabla \tau = 0 \,.
}
This so-called eikonal approximation works in the limit of large $\tau_{\rm norm} \nu$, or in another word, when the angular separations in consideration are much greater than $\theta_{\rm F}$. 
In this limit, the theoretical framework developed in gravitational lensing applies, and we have the useful concepts of image, image magnification, and caustics. 
The geometrical lensing framework has been introduced to the plasma lensing context by \citet{clegg98, pen12,er18} and \citet{wagner20}. 
Compared to the geometric limit of gravitational lensing, the eikonal approximation captures also the 1st order wave effect i.e. interference among different images.
These concepts provide the basic theoretical framework for understanding scintillation phenomena. 
Thus, we introduce them here and in Section.\;\ref{sec:characteristics}.

From the stationary phase condition we can immediately derive the lens equation -- the mapping between the source plane coordinate $\vek{\beta}$ and the image/lens plane coordinate $\vek{\theta}$ as
\eq{
\vek{\theta} - \vek{\beta} = \vek{\alpha} \,,
}
with the deflection angle $\vek{\alpha}(\vek{\theta}) \equiv \nabla \psi(\vek{\theta})$ \citep{schneider92}. The sign of the deflection angle is chosen to be consistent with that defined in the gravitational lensing literature. For a positive $\vek{\theta}$, a converging lens e.g. a gravitational lens leads to a positive $\vek{\alpha}$, and a diverging lens e.g. an overdense plasma lens leads to a negative $\vek{\alpha}$.

Since lensing conserves surface brightness, the magnification $\mu$ of an image is given by the ratio of the surface areas after and before the lensing, i.e. the Jacobian of the lens mapping
\eqs{
\mathcal{A} &= \frac{\partial \vek{\beta}}{\partial \vek{\theta}} = 1 - \partial_{\svek{\theta}} \vek{\alpha} \\
&  = \partial_{\svek{\theta}} \partial_{\svek{\theta}} \tau = 1 - \partial_{\svek{\theta}} \partial_{\svek{\theta}} \psi \,.
} 
The image of an infinitesimally small source at $\vek{\theta}$ is thus brightened or dimmed by a factor $|\mu(\vek{\theta})|$ with
\eq{
\mu = \frac{1}{{\rm det} \mathcal{A}} \,.
}
Lines on the source plane where $\mu$ diverges are called the caustics. 
They separate the regions between which the number of images changes by two.
When a source crosses a caustic, an additional pair of images emerge, or an existing pair of images annihilate.
The infinite magnification at a caustics is a flaw of the eikonal approximation which would not occur in reality even for a point source. 
A full computation with Eq.\;\ref{eq:KF} is required here instead. 

\subsection{Characteristics and Spectral Caustics on Dynamic Spectra}
\label{sec:characteristics}
In gravitational lensing, the observable is usually the 2D image plane. 
In plasma lensing, the source is typically too small to resolve, and the data we have is the light curve of the source. This is equivalent to a 1D slice of the observer's plane, but in addition, we have the frequency dimension.

The frequency dependence encodes a lot of information about pulsar scintillation.
Away from singularities and when the geometric delay dominates, the frequency-axis in a dynamic spectrum can be viewed as a distance indicator, since it is the combination of $\nu \tau \propto \nu D$ that determines the interference pattern. 
In this aspect, a dynamic spectrum is akin to a hologram on a 2D plane that is perpendicular to the image plane.

Singularities of lens mapping appear in a dynamic spectrum as spectral caustics.
As the deflection potential $\psi$ depends on frequency $\nu$, so does the deflection angle:
\eq{
\label{eq:alpha_nu}
\alpha(\vek{\theta}, \nu) = \frac{\nu_0^2}{\nu^2} \alpha_0(\vek{\theta}, \nu_0) \,,
}
plasma lensing caustics forms a two-dimensional surface in the three-dimensional $(\vek{\beta}, \nu)$ space. 
Generically, for each $\vek{\beta}$, there exists some $\nu$ that lies on the caustics, i.e.
with
\eq{
\label{eq:causstics}   
{\rm det}\br{1 - \frac{\nu_0^2}{\nu^2} \partial_{\svek{\theta}} \vek{\alpha_0}} = 0 \,.
}

The transformation Eq.\;\ref{eq:alpha_nu} also defines characteristic lines on the dynamic spectrum:
\eq{
\label{eq:characteristics}
\beta(\vek{\theta}, \nu) = \vek{\theta} - \frac{\nu_0^2}{\nu^2} \alpha_0(\vek{\theta}, \nu_0) \,,
}
as has been pointed out by the pioneering work of \citet{tuntsov16}.
Each characteristic is identified by a common location $\vek{\theta}$ on the lens plane.
Characteristics are useful features to identify on a dynamic spectrum, since they allow us to determine the deflection angle:  
along a characteristic
\eq{
\label{eq:constraint1}
\nu_0^2 \alpha_0 = - \frac{\dd \beta}{\dd \nu^{-2}}  \qquad \qquad \qquad   \rm{\bb{constraint\ I}}\,.
}

In general, a spectral caustic is not a characteristic but an envelope of many characteristics. 
However, by definition, a spectral caustic is always tangent to some characteristic on a dynamic spectrum and thus they share the same local first derivative.
Thus, Eq.\;\ref{eq:constraint1} also applies to points on a spectral caustic which are sometimes easier to identify on dynamic spectra. 
For single-variable lenses which focuses/defocuses only in one direction and is flat in the other direction, the derivative of the deflection angle with respect to the lens plane coordinate $\dot{\alpha} \equiv \partial \alpha / \partial \theta$ is also known along a spectral caustic: 
\eq{
\label{eq:constraint2}
 \frac{\nu_0^2}{\nu^2} \dot{\alpha_0} = 1   \qquad \qquad \qquad \qquad \rm{\bb{constraint\ II}}
}
by definition (cf. Eq.\;\ref{eq:causstics}). 

Therefore, with this particular lens geometry, we can recover $\alpha$ and $\dot{\alpha}$  at each point on a spectral caustic.
The correctness of the assumed lens symmetry can be tested using the magnification along these characteristics away from the caustics at higher frequency when the latter can be identified.
E.g. for a single-variable lens in the cartesian coordinate,
\eq{
\label{eq:constraint3}
\mu(\beta, \nu) = \frac{1}{1 - \dot{\alpha} } = \frac{1}{1 - \br{\nu_0^2 / \nu^2} \dot{\alpha}_0 } \quad \rm{\bb{constraint\ III}}\,.
}
Equations\;\ref{eq:constraint1}, \ref{eq:constraint2} and \ref{eq:constraint3} form three ways to constrain $\alpha$ and $\dot{\alpha}$ that can be applied to dynamic spectra. 
We shall explain their usage in Section.\;\ref{sec:characteristics2lens} (see Table.\;\ref{tab:methods} for a summary).

\section{Simulating dynamic spectra during plasma microlensing}
\label{sec:sim}
\begin{figure}
    \centering
        \includegraphics[width=0.49\textwidth]{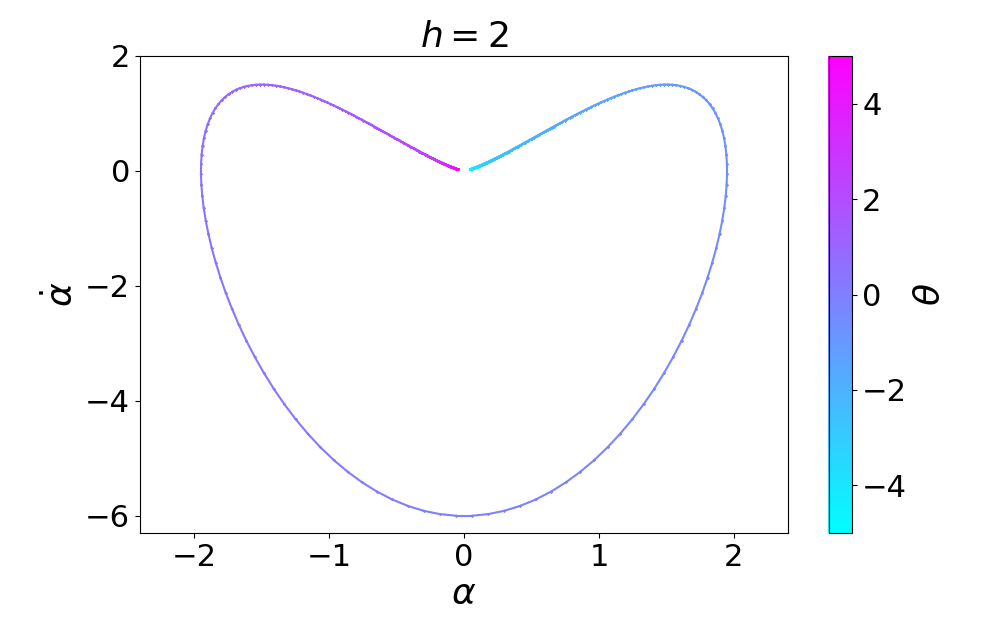}\\
        \includegraphics[width=0.49\textwidth]{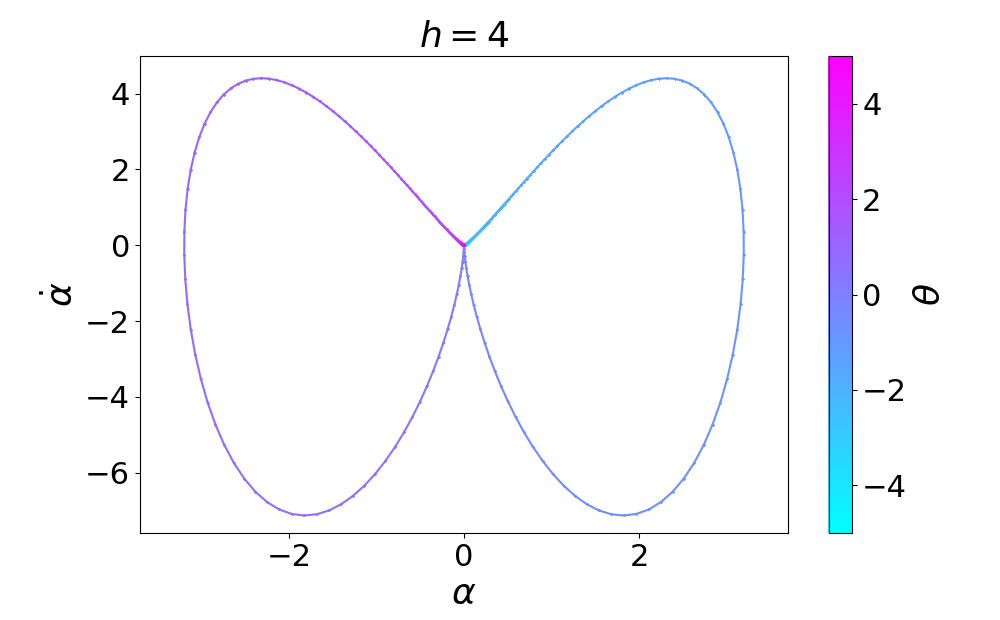}
        \caption{The phase-space diagram of the deflection angle $\alpha$ against its derivative $\partial_{\theta} \alpha$ characterizes the property of a single-variable lens. 
        Upper panel: cored power-law lens (Eq.\;\ref{eq:rational_lens}) with $h=2$; Lower panel: same with $h=4$. The lines are color coded with lens-plane angular coordinate $\theta$. 
        All angles are given in unit of the characteristic width of the lens $\sigma_{\rm lens}$. The amplitue of the lens is chosen to be the default value $A = 3$. The deflection angle for a positive image position $\theta$ is negative, suggesting that the light is originally from a source position $\beta = \theta - \alpha > \theta$. This is typical for the diverging lens corresponding to an overdense patch of plasma. For the same reason, $\dot{\alpha} < 0$ and thus an image is de-magnified $\mu = 1 / (1 - \dot{\alpha}) < 1$ when its angular position is close to the center of the lens.}
    \label{fig:alphaalphadot}
\end{figure}

\begin{figure*}
    \centering
    \includegraphics[width=0.95 \textwidth]{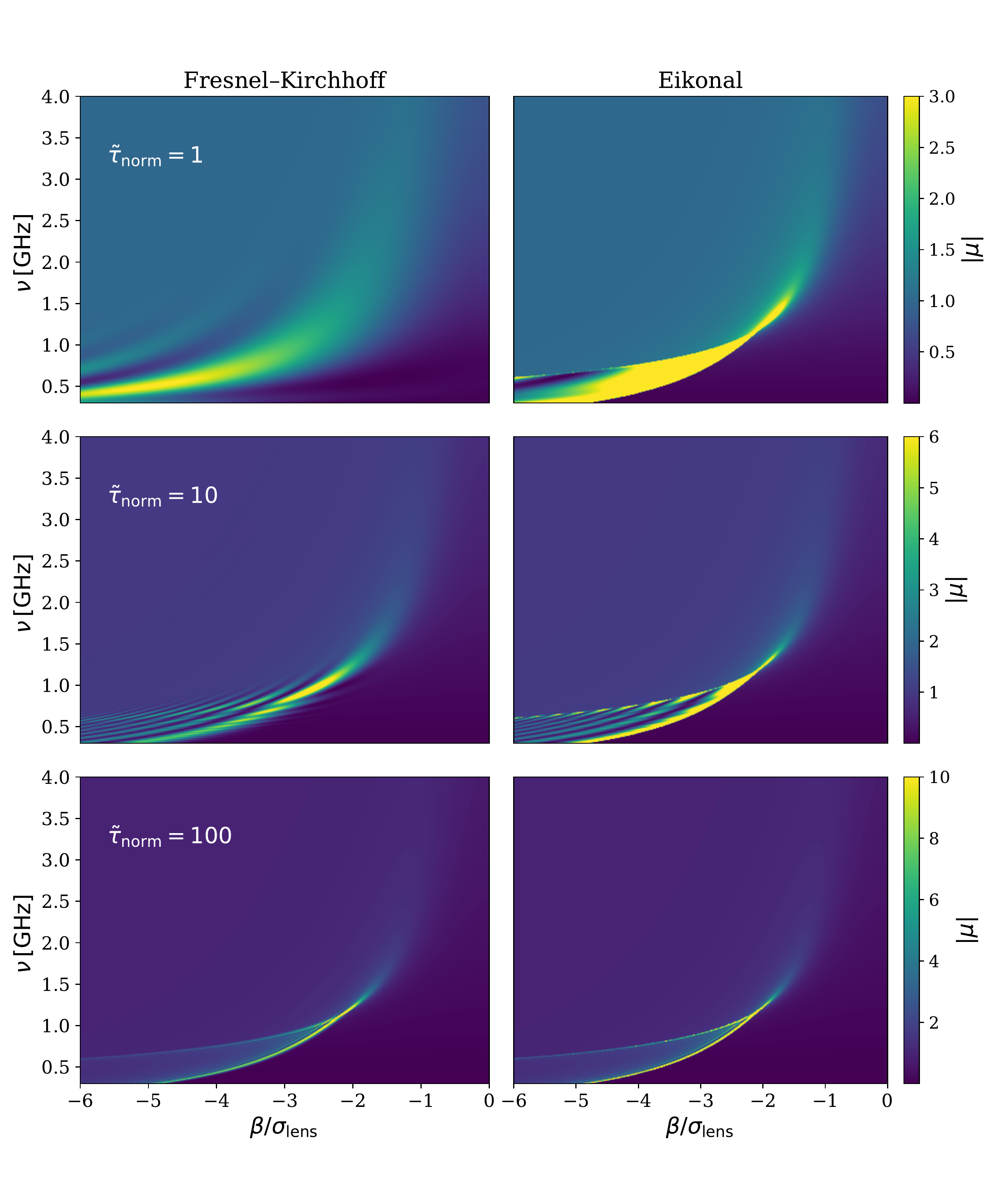}
    \caption{Plasma microlensing dynamic spectra resulting from single-variable cored power-law lens (Eq.\;\ref{eq:rational_lens}) with $h=2$, $A_0 = 3$, and $\nu_0 = 1$ GHz. 
    We compare computation results of the Fresnel-Kirchhoff integral using the Picard-Liftschez method (left panels) and those with the eikonal approximation (right panels). We use the source position ${\beta}$ as a representative of time $t = - \beta  D (1 - f_{\rm d})/ V_{\rm sr}$.
    The simulated dynamic spectra are truncated at $\beta = 0$, i.e. when the source lies right behind the center of the lens.  
    The color indicates the magnification of the flux. 
    Different distance normalizations ($\tilde{\tau}_{\rm norm} = 1$, 10 and 100 GHz$^{-1}$ for the upper, middle, and lower panels, respectively) that correspond to different Fresnel scales $(\theta_{\rm F} / \sigma_{\rm lens})^2 = 1 / (\tilde{\tau}_{\rm norm} \nu)$ have been considered.
    Increasing normalization $\tilde{\tau}_{\rm norm}$ leads to finer fringes, sharper spectral caustics, and higher maximum magnification (note the different upper limits adopted for the color-coding in panels with different distance normalizations). 
    Note that $\tilde{\tau}_{\rm norm} / (2 \uppi)$ in unit of nano second also indicates the typical time delay of the refracted light relative to the undiverted light from the source.
    On dynamic spectra computed with the eikonal approximation (right panels), one can clearly identify the outer and inner spectral caustics as the boundaries of the region with interference patterns where the magnification diverges, and the cusp point as where the two caustics meet (at around $|\beta_{\rm cusp}| / \sigma_{\rm lens} = 2$ and $\nu_{\rm cusp} = 1.2$). 
    The depletion zone can be identified as the region with very low magnifications inside the inner caustics.
    Light curves extracted from these dynamic spectra are presented in Appendix.\;\ref{app:light_curve}.}%
    \label{fig:dyn}
\end{figure*} 

\begin{figure}
    \centering
        \includegraphics[width=0.5\textwidth]{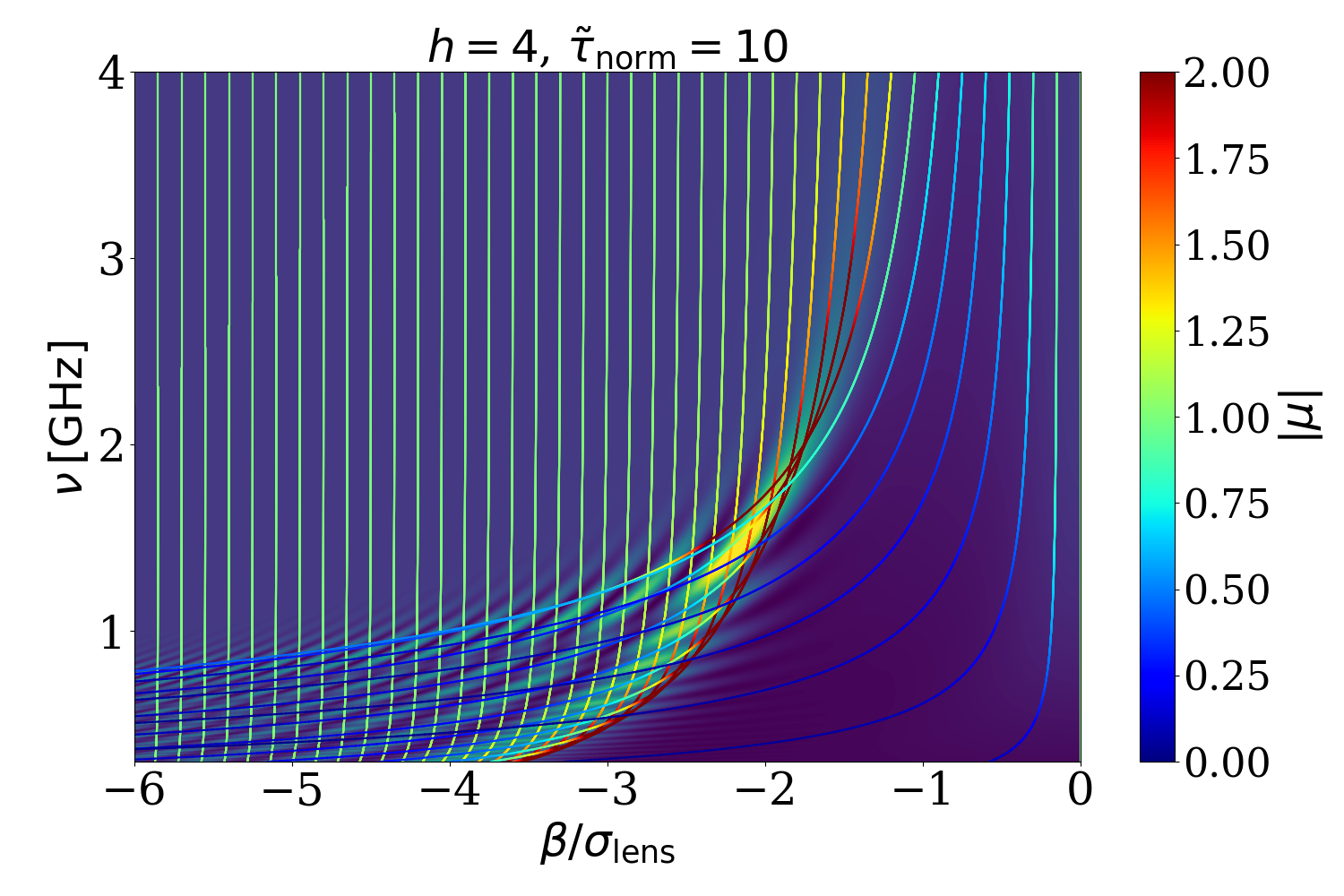}\\
        \includegraphics[width=0.5\textwidth]{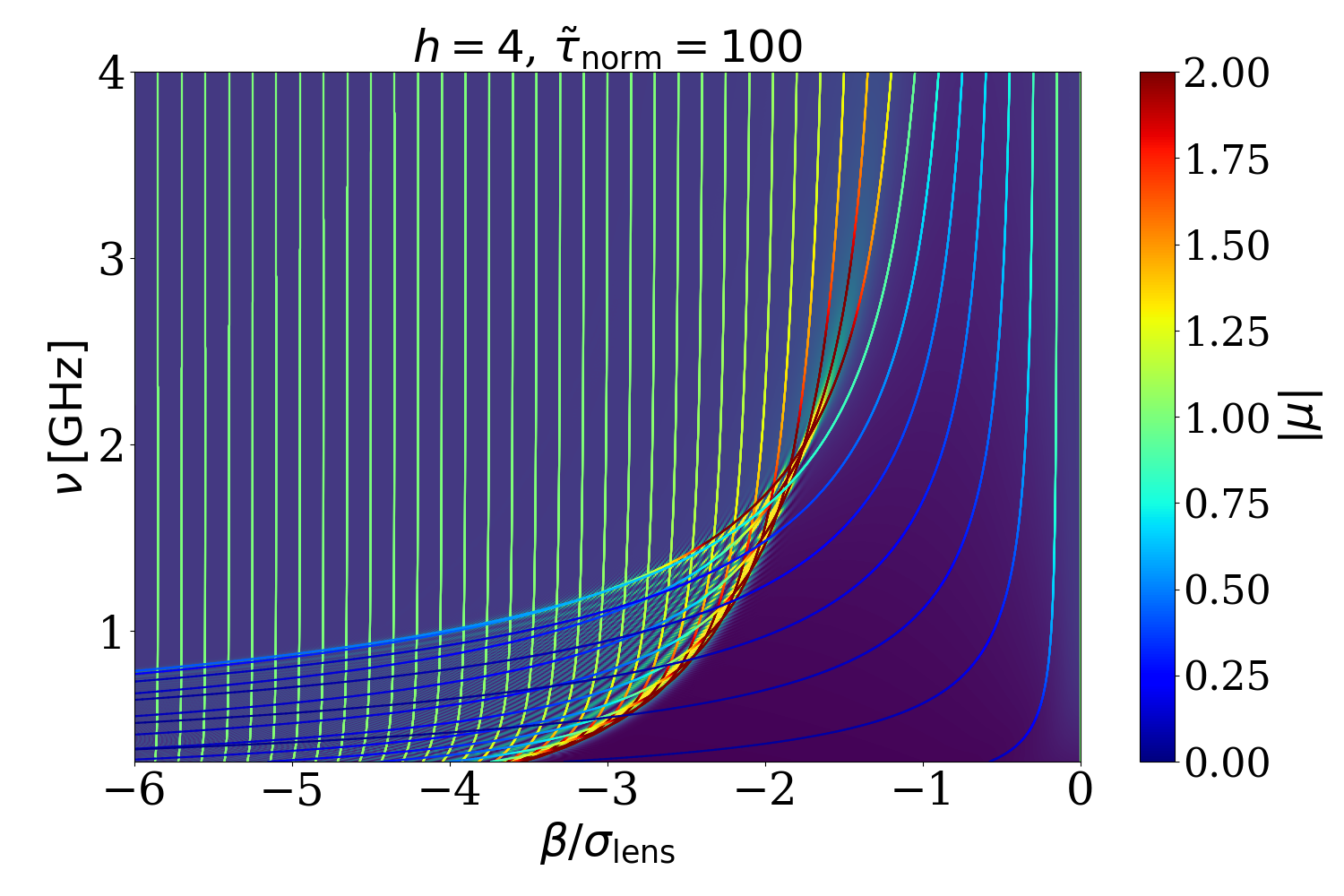}
        \caption{Characteristics overlaid on dynamic spectra for distance normalization $\tilde{\tau}_{\rm norm} = 10$ and 100 GHz$^{-1}$ for the $h=4$ lens. 
        Each characteristic represents one image at a fixed angle $\theta$ from the center of the lens. 
        At different observing frequencies, this image is produced when the source is located at different $\beta$ locations as a result of frequency-dependent deflection angles. 
        The plotted characteristics are selected to space evenly on the $\beta$-axis at $\nu = 4$ GHz. 
        Those having large deflection angles bend more significantly at lower frequencies (Eq.\;\ref{eq:alpha_nu}).
        The color of the characteristics indicates the magnification value computed in the eikonal limit, while 
        the color of the underlying dynamic spectra indicates the actual magnification computed by integrating the Fresnel-Kirchhoff integral. 
        The dynamic spectra use a color-coding same as the middle panels of Fig.\;\ref{fig:dyn}, i.e. with yellow color standing for high magnifications truncated at $|\mu| = 6$.  
        }
        \label{fig:characteristics_on_dynspec}
\end{figure}

\subsection{Lens Models} 
Motivated by the observed properties of the arclets and the \citet{pen14b} model, we consider single-variable lenses that (de-)focus only along one out of the two cartesian coordinates, and a family of lensing potentials in the form of
\eq{
\label{eq:rational_lens}
\psi(\theta) = \frac{A \sigma_{\rm lens}^2}{1 + \br{\theta / \sigma_{\rm lens}}^h} \,,
}
where $\sigma_{\rm lens}$ is the characteristic width of the lens, and $\theta$ is the component of $\vek{\theta}$ in the direction of $\vek{V}_{\rm eff}$. 
For plasma lenses, the normalization is frequency dependent, $A = A_0 \br{\nu /\nu_0}^{-2}$. As default, we choose $A_0 = 3$ and $\nu_0 = 1$GHz.

We put $\sigma_{\rm lens}^2$ explicitly into the normalization of $\psi$ so that the Fermat potential and thus all lensing behaviors in the eikonal limit are invariant with respect to $\sigma_{\rm lens}$ up to an overall normalization. 
In another word, this enables the transformation
\eq{
\tau \to \tilde{\tau}(\tilde{\vek{\theta}}, \tilde{\vek{\beta}}) = \frac{|\tilde{\vek{\theta}} - \tilde{\vek{\beta}}|^2}{2} - \frac{A}{1 + \tilde{{\theta}^h}} \,,  
}
so that $\tilde{\tau}$ as a function of the transformed coordinates $\tilde{\vek{\theta}} = \vek{\theta} / \sigma_{\rm lens}^2$ and $\tilde{\vek{\beta}} = \vek{\beta} / \sigma_{\rm lens}^2$ no longer depends on the lens width $\sigma_{\rm lens}$,
and that
\eq{
\Delta \Phi = \tilde{\tau}_{\rm norm} \nu  \tilde{\tau} = \br{\frac{\sigma_{\rm lens}}{\theta_{\rm F}}}^2 \tilde{\tau} 
}
where $\tilde{\tau}_{\rm norm} = \tau_{\rm norm} \sigma_{\rm lens}^2$. The scaled quantity $\tilde{\tau}_{\rm norm} / (2\uppi)$ has the dimension of time and indicates the typical time delay of the light caused by the plasma lens.

The chosen shape of the lensing potential is a cored power-law which has been widely used and thoroughly studied also in the context of plasma lensing \citep{er18}. 
It also has the merit of being easily generalizable to the complex plane to enable the use of the Picard-Lefschetz theory \citep{feldbrugge19}. 
In addition, the behavior of cored power-law lenses is qualitatively identical to that of the exponential lens family which includes the commonly adopted Gaussian lens \citep{clegg98, romani87, er18, dong18, grillo18}.
The \citet{pen14b} model, however, implies asymmetric fold lenses which we plan to study in future works.

We examine two typical values of the outer slope $h$ of the lensing potential: $h=2$ and $h=4$. 
They represent two basic classes of centrally-condensed lenses characterized by their distinctive phase diagrams (Fig.\;\ref{fig:alphaalphadot}). 
For the $h=2$ lens, $\dot{\alpha}$ has a minimum at the lens center, monotonically increases to its peak value at a certain radius, and then decreases and approaches zero at infinity. 
The $h=4$ lens is a representative of all lenses with $h > 2$. For them, $\dot{\alpha} = 0$ at the lens center, forming an additional local maximum. 
As a consequence, the $h=2$ lens leads to a U-shaped light curve, and the $h=4$ lens a W-shaped light curve (see Appendix.\;\ref{app:light_curve}), both have been observed in ESEs.

\subsection{Numerical Method for the Kirchhoff-Fresnel Integral}
Since we are studying the interference pattern at and around spectral caustics, the eikonal approximation fails to be precise, and direct integration of the Kirchhoff-Fresnel integral (Eq.\;\ref{eq:KF}) has to be performed. 

The Kirchhoff-Fresnel integral is a typical oscillatory path integral that is notoriously hard to compute. 
Early works either consider idealized models \citep{melrose06} or use approximate method \citep{watson06, grillo18} that is valid only near catastrophes. 
However, thanks to a recent breakthrough \citep{feldbrugge19}, a new method that is precise, robust, quickly converging, and at the same time computationally tractable is now available.
This new method generalizes the exponent of the integrand to the complex plane and exploits Cauchy's theorem to transform the oscillatory integral on the real axis into one in the complex plane.
Then it uses the Picard-Lefschetz theory which studies the topology of holomorphic functions to find the saddle points and the `Lefschetz thimbles' -- steepest descent contours connecting them, with each Lefschetz thimble corresponding to a real or an imaginary image \citep{jow21}. 
The relevant Lefschetz thimbles corresponding to the initial integration domain on the real axis can be found by `flowing' the latter into the complex plane along the downward flow of the real part of the exponent, according to the Morse-Smale theory as well as the topology of the holomorphic function.
Integral along these Lefschetz thimbles are then rapidly convergent, and exactly equivalent to the original oscillatory integral. 

We use the \citet{feldbrugge19} method to model the dynamic spectra during plasma microlensing. 
As a comparison, we also compute dynamic spectra under the eikonal approximation. 
The results of the latter would show the spectral caustics clearly and are thus instructive for understanding the features on the dynamic spectra.

\subsection{Results: Eikonal vs Kirchhoff-Fresnel}
\label{sec:sim_res}
Fig.\;\ref{fig:dyn} presents the simulated dynamic spectra for the $h=2$ lens for three different values of the distance normalization $\tilde{\tau}_{\rm norm} = 1, 10$ and 100 GHz$^{-1}$. These correspond to $\sigma_{\rm lens}/\theta_{\rm F} \approx 1, 3 $ and $10$ at 1 GHz, respectively. If we take $D = 1$ kpc and $f_{\rm d} = 0.5$, the width of the plasma lens would be $\sigma_{\rm lens} \approx 0.01, 0.03$ and 0.1 mas for these distance normalizations (see Eq.\;\ref{eq:tau_norm}), and the Fresnel scale $\theta_{\rm F} \approx 0.01 (\nu /\rm GHz)^{-1/2}$ mas. 
We have simulated the dynamic spectra spanning a wide frequency band (0.4 - 4 GHz) and a duration corresponding to $-6 < \beta / \sigma_{\rm lens} < 0$, with $\beta = 0$ being the time of perfect source-lens alignment. 
Dynamic spectra over this wide range allow us a complete view of the features left by a plasma microlensing event.
The $h=4$ lens gives dynamic spectra that are qualitatively the same as those by the $h=2$ lens for most features we are going to discuss. 
In Fig.\;\ref{fig:characteristics_on_dynspec}, we show the $h=4$ lens dynamic spectra simulated with the Kirchhoff-Fresnel integral for $\tilde{\tau}_{\rm norm} = 10$ and 100 GHz$^{-1}$. 
We have overlaid some characteristics given by the transformation Eq.\;\ref{eq:characteristics}. 
The color of the characteristics shows the local magnification value in the eikonal limit.

The eikonal approximation (right panels of Fig.\;\ref{fig:dyn}), although it fails to represent the actual dynamic spectra as the distance normalization decreases (i.e. $\sigma_{\rm lens}$ approaches $\theta_{\rm F}$), 
is very helpful for identifying the key features in the dynamic spectra: the cusp point, the inner and outer spectral caustics, and the regions divided by them.  

The cusp point can be identified as the brightest location on the dynamic spectra in the eikonal approximation at around $|\beta_{\rm cusp} / \sigma_{\rm lens}| = 2$ and $\nu_{\rm cusp} = 1.2$ in the right panels of Fig.\;\ref{fig:dyn}.
For light with frequency $\nu > \nu_{\rm cusp}$, the lens is not strong enough to focus it at the observer plane.
Thus, at frequencies above $\nu_{\rm cusp}$, the lens mapping is non-critical ($\rm{det}\mathcal{A}$ is nowhere zero), and the lens produces a single image at all times. 
Close to source-lens alignment, there is a region where the image is de-magnified ($|\mu| < 1$) as a result of the light-diverging property of plasma overdensities. 
The light rays diverted from the center of the lens concentrate in a region along the cusp characteristic i.e. the characteristic that passes through the cusp point (Fig.\;\ref{fig:characteristics_on_dynspec}, see also Sect.\;\ref{sec:cc}), causing the image to be magnified in this narrow region.
At large $|\beta|$ away from this magnified region, the image is little affected by the lens, $\nu \approx 1$.

Below the cusp frequency $\nu < \nu_{\rm cusp}$, the lens mapping is critical at two $|\beta|$'s at each frequency, forming two spectral caustics on each side of source-lens alignment (right panels of Fig.\;\ref{fig:dyn}).
These two spectral caustics divide regions with different image multiplicity: the lens produces three images in the region surrounded by them, and one image outside.  
In the three-image zone, the interference of the images leads to interference patterns. 
Inside the inner caustic (the one closer to $\beta = 0$), there exists a depletion zone with the same origin as the de-magnified region at $\nu > \nu_{\rm cusp}$, 
but wider, more de-magnified, and with sharper boundary than the de-magnified region.
Outside the outer caustic, the image is again not much affected by the lens.

At large distance normalizations $\tilde{\tau}_{\rm norm} \gtrsim 10$ GHz$^{-1}$ i.e. $\sigma_{\rm lens} \gg \theta_{\rm F}$, it is possible to locate the above features on an actual dynamic spectrum (left panels of Fig.\;\ref{fig:dyn}). 
As $\tilde{\tau}_{\rm norm}$ decreases, the brightest location moves away from the cusp point, and the spectral caustics fade as the neighboring rays become more coherent in their phases: 
With $\tilde{\tau}_{\rm norm} \sim 1$ GHz$^{-1}$, the Fresnel scale $\theta_{\rm F}$ becomes as large as the length scale $\sigma_{\rm lens}$ on the dynamic spectra for the $\nu \sim 1$ GHz frequency range we are probing.
This would occur in reality only if the lens size is very small, with $\sigma_{\rm lens}$ in terms of physical size being less than 0.01 AU.  
Larger, AU sized lenses would correspond to even greater $\tilde{\tau}_{\rm norm}$ values than the maximum one we consider here ($1\;\rm{AU} / 1\;\rm{kpc} \approx 1 \; \rm{mas}$). 
Thus, for them, the patterns on the microlensing dynamic spectra would be well-captured by the computations in the eikonal limit.

\section{Constraining lens properties}
\label{sec:constrain}
Now that we have simulated the observable of a plasma microlensing event -- the dynamic spectrum, we move on to consider how can one infer the properties of the plasma lens  
with it. 

\subsection{Cusp Point/Characteristic to Lens Size and Amplitude}
In gravitational lensing, the radius of the Einstein ring offers a direct measure of the mass scale of the lens when the source and the lens are perfectly aligned. 
Considering an axisymmetric lens, the Einstein ring corresponds to a tangential critical curve within which the mass is just equal to the critical mass to focus the light at the observer. 
In a centrally condensed, diverging plasma lens, there exists no tangential critical curve \citep[e.g.][]{er18, er19} and thus no correspondence of the Einstein ring. 
Nevertheless, the locations of the features on a plasma microlensing dynamic spectrum offer clues to the size and strength of the plasma lens.

\subsubsection{Cusp point}
We take the cusp point, the feature that is the easiest to identify if captured observationally, as an example. 
The lens amplitude at the cusp frequency $A_{\rm cusp} = A_0 (\nu_{\rm cusp} / \nu_0)^{-2}$ corresponds to the minimum amplitude to make the lens critical, 
whereas the temporal location of the cusp point $\beta_{\rm cusp}$ gives a typical size of the depletion zone. 
They can be computed precisely given a lens model and depends on the model parameter, but as an order of magnitude estimate, $|\beta_{\rm cusp}| \gtrsim \sigma_{\rm lens}$ for all $h$ values, and $A_{\rm cusp} \sim 1$ for moderate $h$ values (see Appendix.\;\ref{app:cusp} for details).  

These estimations at the cusp point can be combined to constrain the column density of the plasma lens. 
Combining the physical expression of the deflection potential (Eq.\;\ref{eq:psi_phys}) and the model (Eq.\;\ref{eq:rational_lens}), we have
\eqs{
\label{eq:Ne_est}
 & \frac{N_{\rm e}}{\rm cm^{-3} \times AU}  =   \frac{120 A_{\rm cusp}}{1 + f(h)} \br{\frac{\sigma_{\rm lens}}{\rm mas}}^2 \br{\frac{\nu_{\rm cusp}}{\rm GHz}}^{2} \br{\frac{D}{\rm kpc}} \br{\frac{1 - f_{\rm d}}{f_{\rm d}}}   \\ 
& \approx  \frac{7 \times 10^{-4} A_{\rm cusp}}{1 + f(h)} \br{\frac{|t_{\rm cusp}|}{\rm hour}}^2 \br{\frac{\nu_{\rm cusp}}{\rm GHz}}^{2} \br{\frac{V_{\rm sr}}{100\,\rm km/s}}^2 \br{\frac{D}{\rm kpc}}^{-1} \frac{1}{f_{\rm d} (1 - f_{\rm d})}
}
with $f(h)$ being a lens-shape dependent factor. 
For our lens model, $f(h) \equiv \br{\theta_{\rm cusp} / \sigma_{\rm lens}}^h \approx 1$ for $h \gtrsim 2$. We have used $|\beta_{\rm cusp}| \approx \sigma_{\rm lens}$ in the approximation on the second line, and have converted $\beta$ into the actual observable $t$.
The combination $V_{\rm sr}^2 D^{-1}f_{\rm d}^{-1} (1 - f_{\rm d})^{-1}$, if not known a priori, can be determined from the width of the interference pattern (see Section.\;\ref{sec:interference}).

\subsubsection{Cusp characteristic}
\label{sec:cc}
In reality, one needs to be lucky to capture the cusp point in an observation given limited time and frequency sampling. 
As an alternative, one could exploit the maximumly magnified characteristic at $\nu > \nu_{\rm cusp}$ which is much easier to capture observationally. This characteristic forms the bright ridge in the dynamic spectrum extending from the cusp point to higher frequencies, and thus we refer to it as the `cusp characteristic'. Measuring the cusp characteristic location $\beta_{\rm cc}$ and magnification $\mu_{\rm cc}$ as functions of observing frequency can enable the determination of the location of the cusp point $\beta_{\rm cusp}$ and $\nu_{\rm cusp}$ (Appendix.\;\ref{app:cusp}).


\subsection{Caustics / Characteristics to Lens Shape}
\label{sec:characteristics2lens}

\begin{figure}
    \centering
        \includegraphics[width=0.49\textwidth]{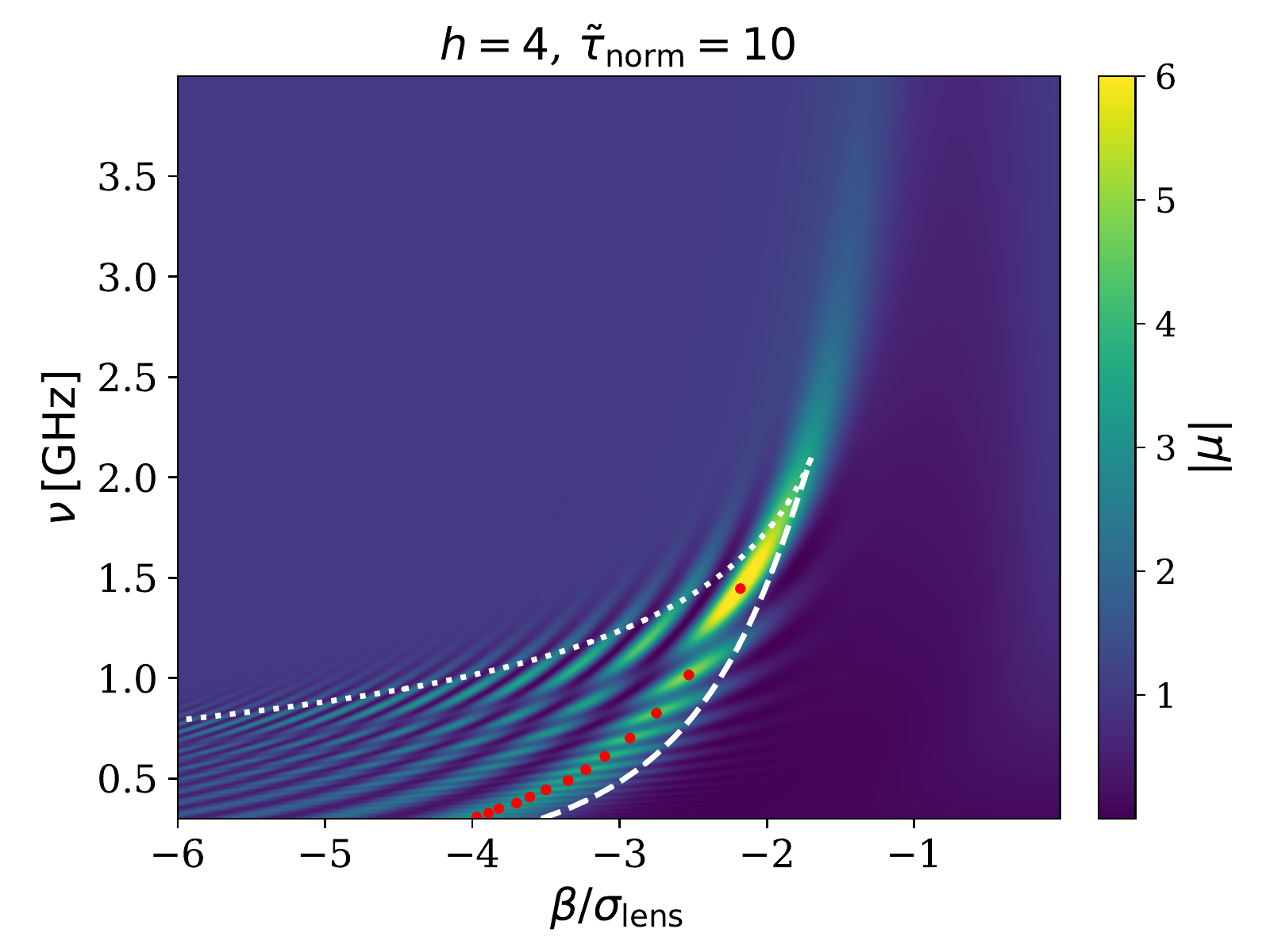}\\ 
        \includegraphics[width=0.49\textwidth]{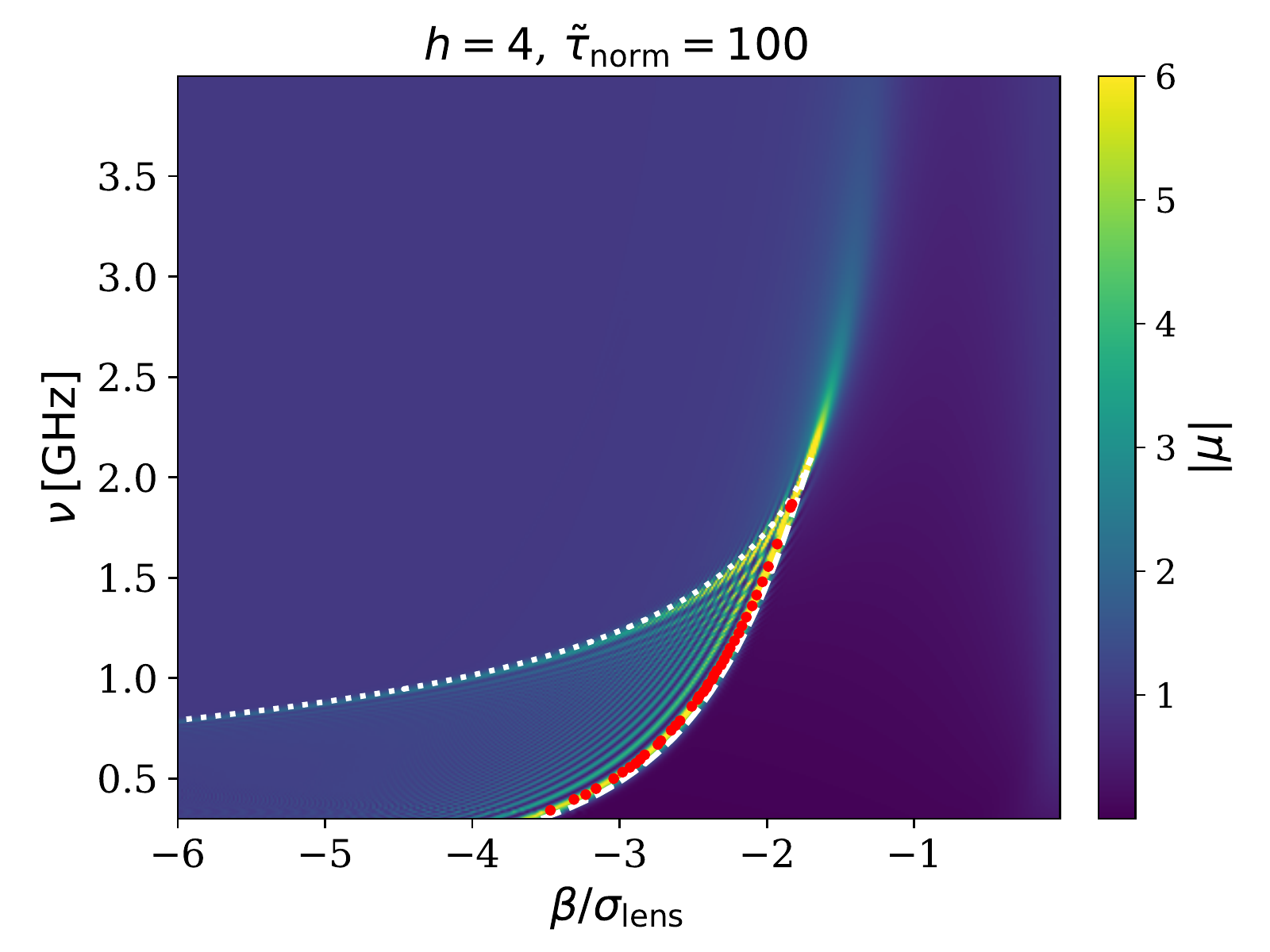}
        \caption{Locations of the actual inner caustic (white dashed line) and outer caustic (white dotted line) overlaid on the microlensing dynamic spectra for cored power-law lenses with $h=4$. 
        Local maxima of the dynamic spectrum on the boundary of the depletion zone (red points) can be used as an approximation for the location of the inner caustics for $\tilde{\tau}_{\rm norm} \sim 10$ GHz$^{-1}$ and above. 
        They can be used to constrain the outer slope of the plasma lens, see Figs.\;\ref{fig:caustics2lens} and \ref{fig:caustics2outerslope} and Section.\;\ref{sec:characteristics2lens} for details. }
        \label{fig:caustics_est}
\end{figure}

\begin{figure}
\centering
    \includegraphics[width=0.47\textwidth]{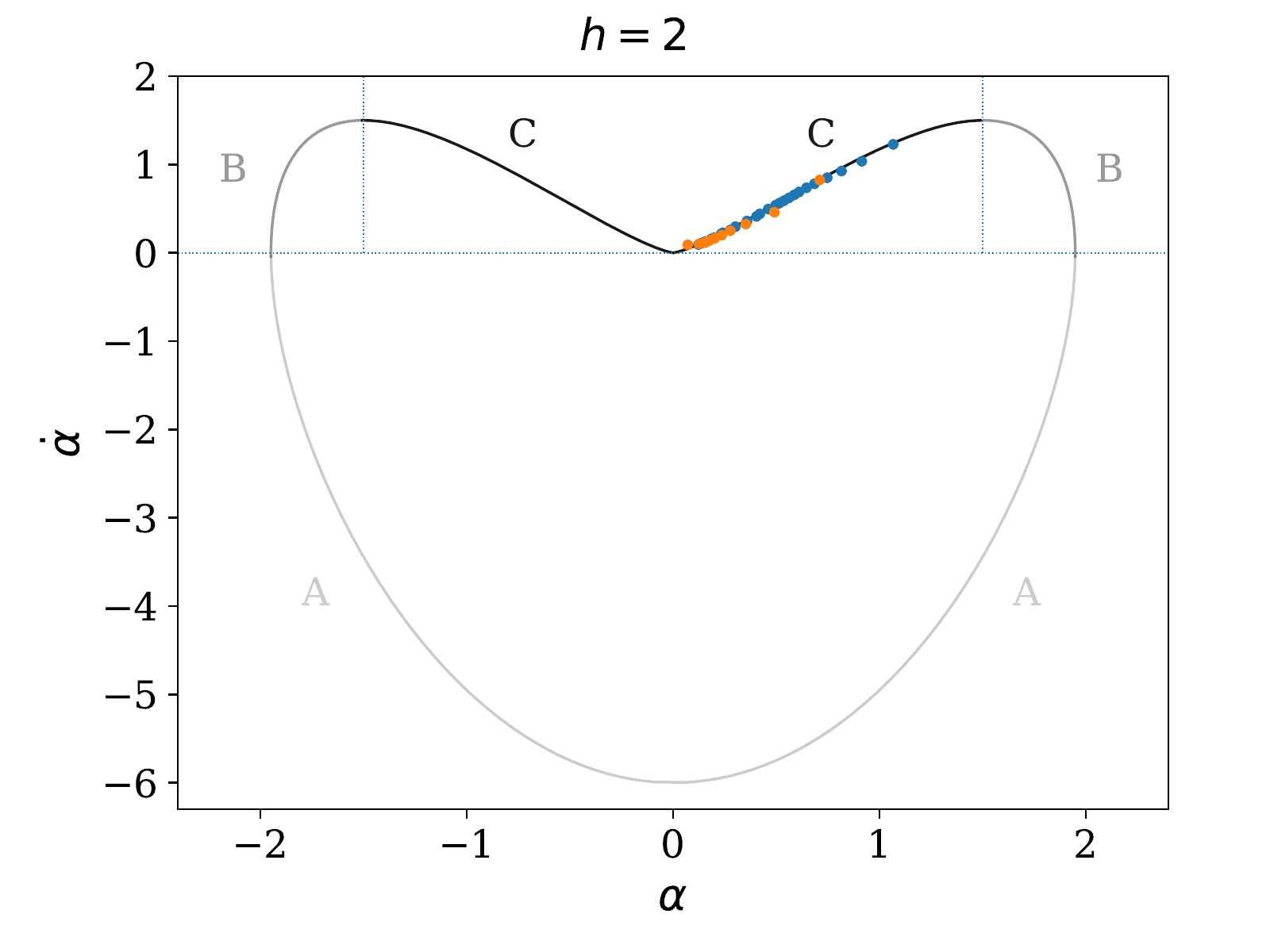}\\
    \includegraphics[width=0.47\textwidth]{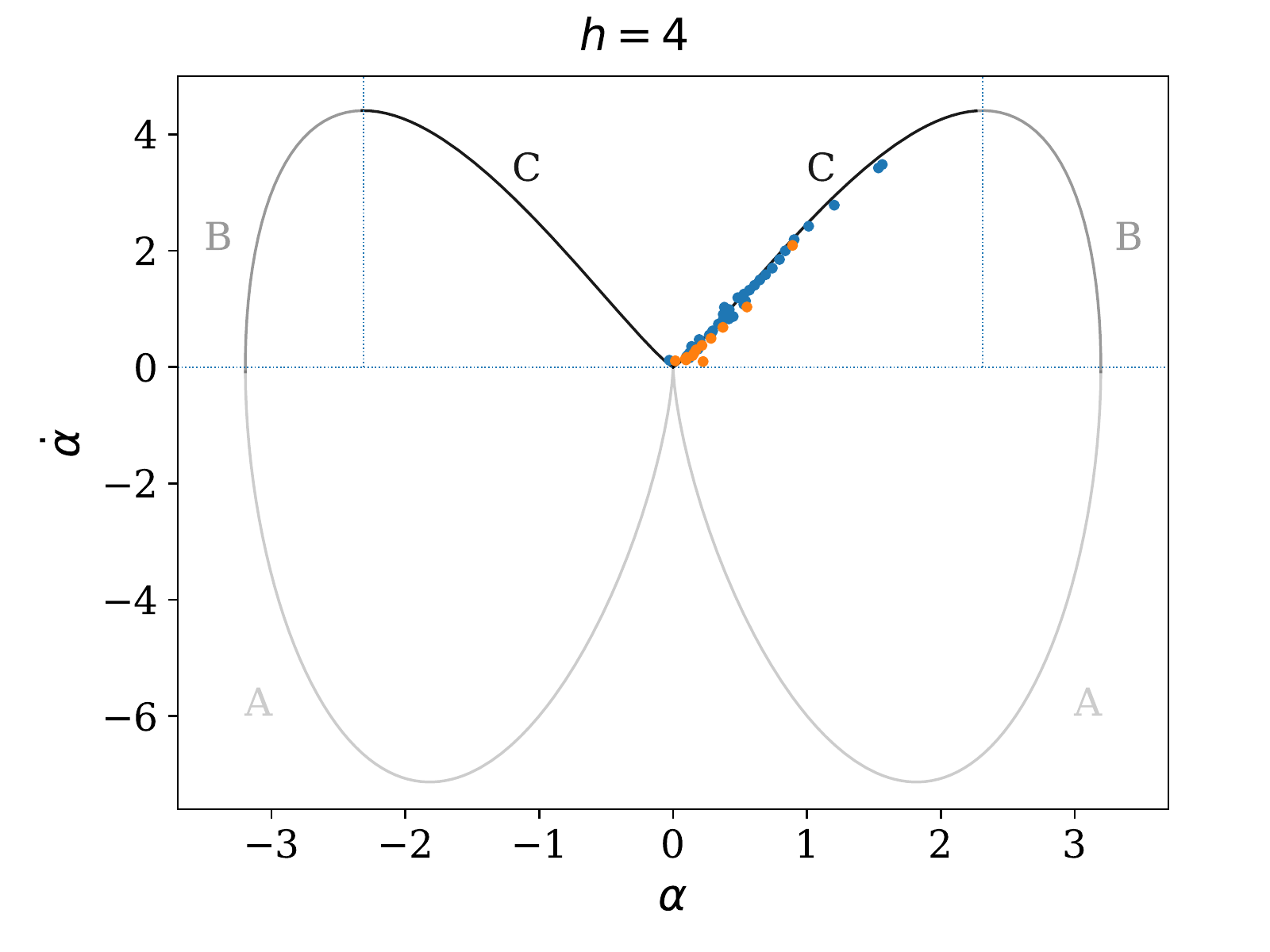}
    \caption{Deflection angle $\alpha$ and its derivative $\dot{\alpha}$ estimated from the inner caustic in the dynamic spectra can constrain section C of the $\alpha - \dot{\alpha}$ diagram where $\alpha$ is increasing but $\dot{\alpha}$ is decreasing with $|\theta|$. 
    Data points are extracted from the local maxima of the dynamic spectrum just outside of the depletion zone as approximations for the inner caustics (see Fig.\;\ref{fig:caustics_est}) for $\tilde{\tau}_{\rm norm} = 100$ GHz$^{-1}$ (blue points) and $\tilde{\tau}_{\rm norm} = 10$ GHz$^{-1}$ (orange points).
    They are then mapped along the corresponding characteristics to the reference frequency $\nu_0 = 1$ GHz to constrain the lens at a fixed amplitude.
    The deflection angle $\alpha$ and its derivative $\dot{\alpha}$ for the lens at $\nu_0 = 1$ GHz are derived using the frequency-dependence of the caustic location (constraint I, Eq.\;\ref{eq:constraint1}) and the diverging magnification (constraint III, Eq.\;\ref{eq:constraint3}), respectively.
    Sections A and B on the $\alpha - \dot{\alpha}$ diagram can be probed by the depletion region and the outer caustic, respectively (see Section.\;\ref{sec:characteristics2lens} for details and Table.\;\ref{tab:methods} for a summary). 
    Section A $\to$ B $\to$ C corresponds to an increasing $|\theta|$, see Fig.\;\ref{fig:alphaalphadot}. }
    \label{fig:caustics2lens}
\end{figure}

\begin{figure}
\centering
    \includegraphics[width=0.47\textwidth]{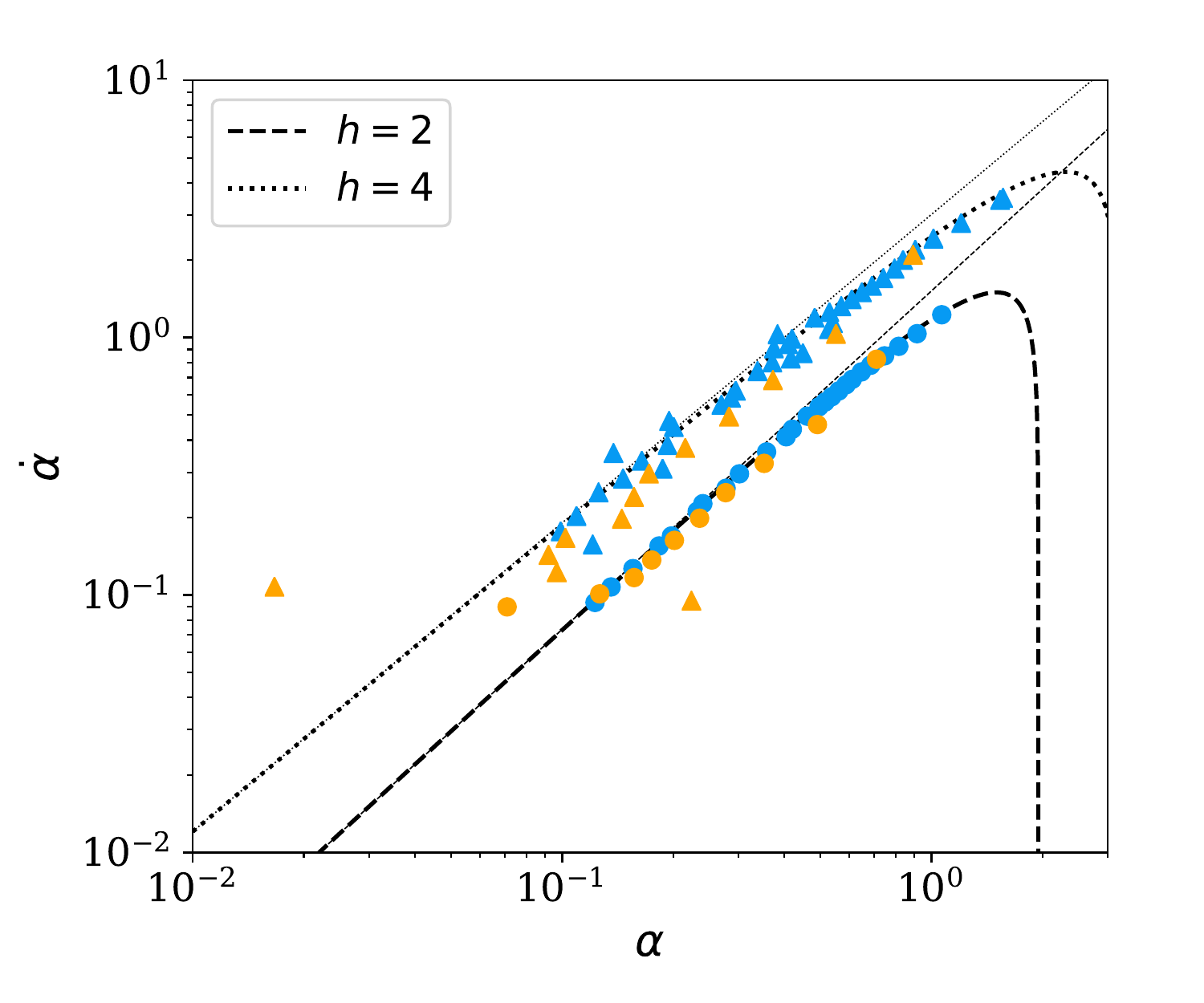}\\
    \caption{The power-law slope of the $\dot{\alpha} - \alpha$ relation derived from the inner caustic can constrain the outer slope $h$ of the lens potential. 
    The data points are the same as those in Fig.\;\ref{fig:caustics2lens}. 
    Thick lines show the actual power-law slopes of the lenses which change with the coordinates, and the thin lines show the asymptotes at large distances from the lens center, whose power-law slopes are $(h+2)/(h+1)$.}
    \label{fig:caustics2outerslope}
\end{figure}

\begin{table*}
\begin{tabular}{|l|l|l|l|}
\rowcolor[HTML]{DAE8FC}
\hline
                                       & Section A ($\dot{\alpha} < 0$)           & Section B ($\dot{\alpha} > 0$, sgn($\theta$)$\ddot{\alpha} > 0$) & Section C ($\dot{\alpha} > 0$, sgn($\theta$)$\ddot{\alpha} < 0$) \\
\hline
\textbf{under-focused} ($\nu > \nu_{\rm cusp}$) region  & II + III in de-magnified regions & II + III in magnified regions  & II + III in magnified regions  \\
\hline
\rowcolor[HTML]{EFEFEF}
\textbf{over-focused} ($\nu < \nu_{\rm cusp}$) region  & II + III in depletion zone \footnotemark   & I + II on outer caustic & I + II on inner caustic \\
\hline
\end{tabular}
\caption{Summary of methods to reconstruct the $\alpha - \dot{\alpha}$ diagram of a plasma lens using plasma microlensing dynamic spectra. 
The $\alpha - \dot{\alpha}$ diagram has been divided into sections A, B and C according to the signs of the deflection angle $\alpha$ and its derivative $\dot{\alpha}$ (see Fig.\;\ref{fig:caustics2lens}).
For different sections, different reconstruction methods apply. 
The method in choice also depends on whether we are using the under-focused or over-focused regions on the dynamic spectrum (see Section.\;\ref{sec:sim_res} for the features on the dynamic spectrum).
Constraints I, II and III are given in Eqs.\;\ref{eq:constraint1}, \ref{eq:constraint2} and \ref{eq:constraint3}, respectively.  
See section.\;\ref{sec:characteristics2lens} for details.}
\label{tab:methods}
\end{table*}
\footnotetext{Section A can also be constrained by analyzing the interference pattern at $\nu < \nu_{\rm cusp}$ in the Fourier domain (Appendix.\;\ref{app:sec}).}

To constrain the shape of the plasma lens, we need the shape of dynamic spectrum features.
In this respect, the shape of the inner spectral caustic is the easiest to obtain observationally. 
Here, we demonstrate constraining the lens shape with the inner caustic and will discuss the use of other features. 

On dynamic spectra with $\tilde{\tau}_{\rm norm} \gtrsim 10$ GHz$^{-1}$, we can approximate the location of the inner caustic with local maxima of the dynamic spectrum near the depletion zone (Fig.\;\ref{fig:caustics_est}). 
We then impose the constraints I (Eq.\;\ref{eq:constraint1}) and II (Eq.\;\ref{eq:constraint2}) which apply to all points on spectral caustics to map these points to the corresponding points at the reference frequency $\nu_0$.
The latter then constrains a part of the $\alpha - \dot{\alpha}$ plane for the plasma lens with $A = A_0$ (Fig.\;\ref{fig:caustics2lens}).
As is shown by Fig.\;\ref{fig:caustics2lens}, the estimates align well with the theoretical values, confirming the validity of local maxima as an estimation for the inner caustic.

Near the origin on the $\alpha - \dot{\alpha}$ plane, these data points have small deflection angles and are produced when the lens is far from being aligned with the source. 
In another word, they correspond to large $|\theta|$ values, see Fig.\;\ref{fig:alphaalphadot}.
Thus, they can in principle constrain the outer logarithmic slope $h$ of the lensing potential, as demonstrated by Fig.\;\ref{fig:caustics2outerslope}.  
The logarithmic slope on the $\alpha - \dot{\alpha}$ plane relates to $h$ as $\dd \ln \dot{\alpha} / \dd \ln \alpha \to (h+2) / (h+1)$ when $\alpha \ll 1$.
For this purpose, however, the rough method we use to extract the inner spectral caustic is not precise enough to convincingly distinguish the slopes at $h = 2$ and 4.

Spectral caustics trace only the increasing sections of $\alpha(\theta)$ since criticality occurs only along characteristics with $\dot{\alpha} > 0$ (sections B and C in Fig.\;\ref{fig:caustics2lens}).
The outer caustic traces the section where $\dot{\alpha}(|\theta|)$ is increasing (section C), and the inner caustic the section where $\dot{\alpha}(|\theta|)$ is decreasing (section B). 
They are separated by the cusp point where $\dot{\alpha}(\theta)$ reaches its maximum.

Away from the caustics, we no longer have the constraint on $\dot{\alpha}(\theta)$ given by the contraint II (Eq.\;\ref{eq:constraint2}). 
However, if we can identify a characteristic, we can use the masurement of $|\mu|$ along it to infer $\dot{\alpha}(\theta)$ with contraint III (Eq.\;\ref{eq:constraint3}).
This constraint, which relies on the eikonal approximation, does not work on the caustics.

For frequencies above that of the cusp ($\nu > \nu_{\rm cusp}$), the shape of the brightness peak in the dynamic spectrum follows the orientation of a characteristic. 
This particular characteristic marks where the light rays are most converged before they reach the focus, which makes it easily identifiable from the dynamic spectrum. 
Combining ${\alpha}(\theta)$ and $\dot{\alpha}(\theta)$ derived from constraint I and III, one can constrain the transition region between sections B and C on the $\alpha - \dot{\alpha}$ diagram using this characteristic.

The method of characteristics is well applicable in the under-focused regime (where $\nu > \nu_{\rm cusp}$) in general. 
The \citet{tuntsov16} work offers a nice example of lens reconstruction in this regime using a combination of constraints II and III. 
Their method works for arbitrarily complicated lens shapes, and all three sections A, B, and C can be constrained given sufficiently good measurements of the flux variations. 

The section of $\alpha - \dot{\alpha}$ where $\alpha(\theta)$ is decreasing (section A) is traced by characteristics in the de-magnified region, since a negative $\dot{\alpha}(\theta)$ means $|\mu| < 1$ (see Eq.\;\ref{eq:constraint2}). In the depletion zone, the de-magnification associated with these characteristics makes it hard to identify them or to measure $|\mu|$ along them, which affects the derivation of the shape of section A from the over-focused regime. 
Nevertheless, important qualitative constraints are possible: The magnification $|\mu|$ at the geometrical center would give us the $\dot{\alpha}$ value at $\alpha=0$. 
It is this very information that distinguishes the $h = 2$ and $h > 2$ families of lenses. 
When the maximum de-magnification does not occur at the geometrical center (when $h > 2$), the value of minimum $|\mu|$ and where it occurs reveal information about the location of the minimum $\dot{\alpha}$ on the $\alpha - \dot{\alpha}$ diagram. Besides constraints using characteristics on a dynamic spectrum, section A could also be constrained by analyzing the three-image zone in the Fourier domain. There the de-magnified images interfere with the bright main image, which makes them easier to measure (Appendix.\;\ref{app:sec}). 

We summarize these methods discussed above in Table.\;\ref{tab:methods}.

\subsection{Interference Pattern to Distance-Velocity Combination}
\label{sec:interference}
\begin{figure}
    \includegraphics[width=0.48 \textwidth]{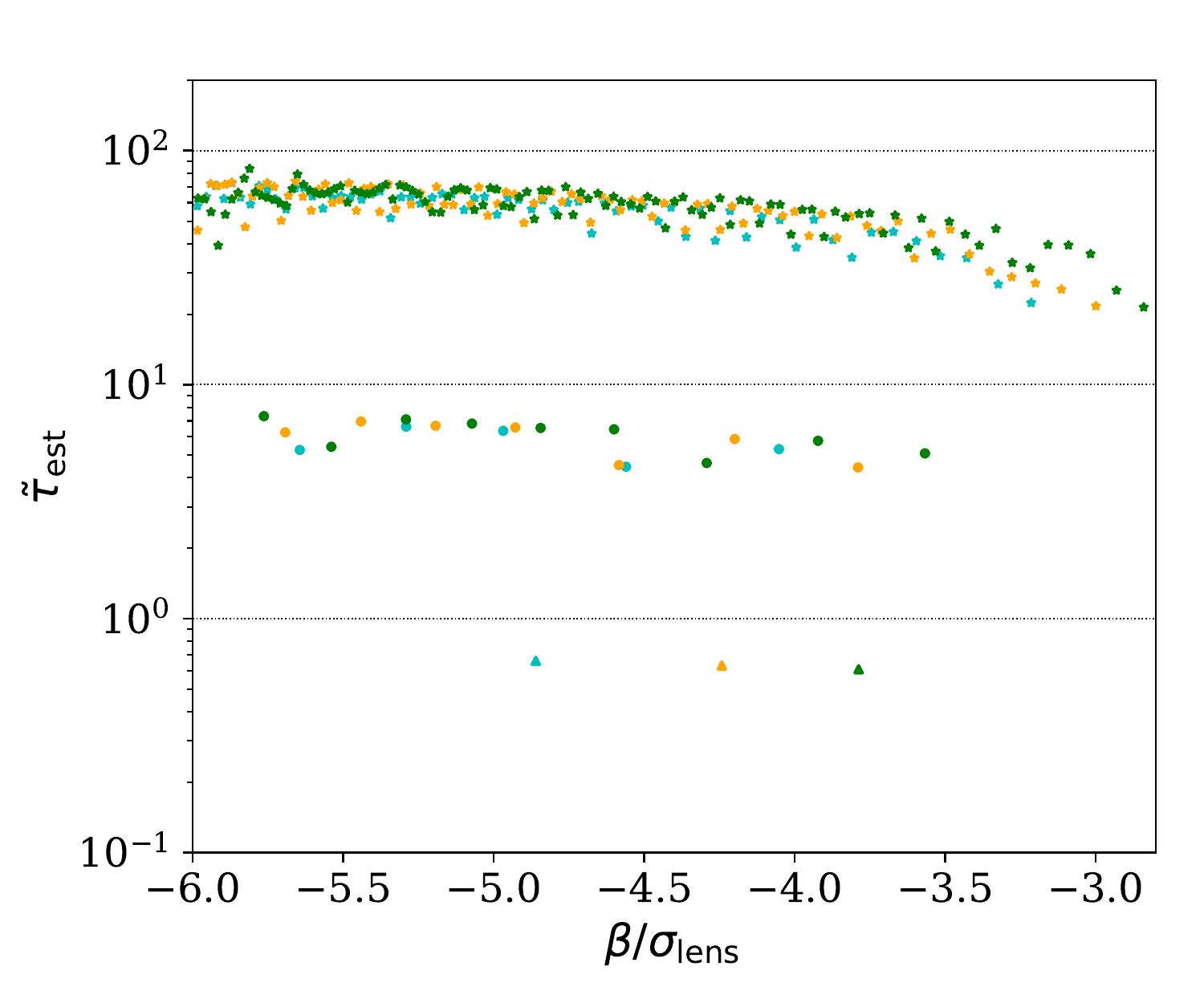}
    \caption{Estimation of $\tilde{\tau}_{\rm norm}$ from the fringe pattern in the dynamic spectrum using Eq.\;\ref{eq:tau_est}. 
    Dynamic spectra for the $h=4$ lens are used.
    Stars, circles and triangles are estimates $\tilde{\tau}_{\rm norm} = $100, 10 and 1 GHz$^{-1}$, respectively. 
    Points are taken from three frequency slices of the dynamic spectrum: $\nu =$ 0.5 (cyan), 0.6 (orange) and 0.7 (green) GHz.
    The actual values of $\tilde{\tau}_{\rm norm}$ are marked as the horizontal lines.}
    \label{fig:fringe2norm}
\end{figure}  

Comparing simulated dynamic spectra at different $\tilde{\tau}_{\rm norm}$ (Figs.\;\ref{fig:dyn} and \ref{fig:characteristics_on_dynspec}), difference in the width of the fringe pattern is the most evident.
The higher $\tilde{\tau}_{\rm norm}$ is, the faster the phases change and the narrower the fringe pattern. 
Here we demonstrate how to use the latter to infer the former.

Take a closer look at the characteristics in the three-image zone (Fig.\;\ref{fig:characteristics_on_dynspec}). 
Recall that one characteristic line corresponds to one image, one can see that among the three images, the most de-magnified image corresponds to a characteristic in the de-magnified region at $\nu > \nu_{\rm cusp}$; another image has $|\mu| \approx 1$ and corresponds to a characteristic that lies almost at fixed $\beta$ when the frequency varies; the third image corresponds to a characteristic near the cusp characteristic at $\nu > \nu_{\rm cusp}$, and is strongly magnified near the inner caustic and de-magnified well within the three-image zone. 
The fringe pattern is dominated by the interference between the latter two images which are the brighter ones among the three.
The phase variation that leads to the interference pattern is further dominated by that from the third image since the second image is produced close to the source $\theta \approx \beta$, and thus its phase changes very little with time.

Assuming the phase variation as being dominated by the geometrical delay of this image, we infer that the phase of this image must change by $2\uppi$ at neighboring fringes, i.e., $\Delta \Phi_{i, i+1} = 2\uppi$ with
\eqs{
    \label{eq:fringe}
    \Delta \Phi_{i, i+1} 
    & \approx \tilde{\tau}_{\rm norm}  \br{\nu \tilde{\beta} \Delta \tilde{\beta}  + \frac{|\tilde{\theta} - \tilde{\beta}|^2}{2} \Delta \nu } \\
    & \approx \tilde{\tau}_{\rm norm} \nu \Delta \tilde{\beta} \br{\tilde{\beta} + \frac{\tilde{\theta} - \tilde{\beta}}{4}} \,.
} 
In the second approximation, we have used $\dd \tilde{\beta} / \dd \nu = 2 (\tilde{\theta} - \tilde{\beta}) / \nu$ derived from the characteristic equation Eq.\;\ref{eq:characteristics}.

As an order-of-magnitude approximation, we neglect the second term on the right-hand side and design an estimator of $\tilde{\tau}_{\rm norm}$ as
\eq{
\label{eq:tau_est}
\tilde{\tau}_{\rm est} = \frac{2\uppi}{\nu \tilde{\beta} \Delta\tilde{\beta}} \,.
}
This allows for an estimate of $\tilde{\tau}_{\rm norm}$ using a single-frequency light curve,
with the location $\tilde{\beta}$ and width $\Delta \tilde{\beta}$ of each fringe taken at certain frequency $\nu$.
We sample the fringe patterns at points with magnification $|\mu| \simeq 1$, and compute the distance between neighboring fringes $\Delta \tilde{\beta}$ as the twice the distance between neighboring sampling points.

Fig.\;\ref{fig:fringe2norm} presents a comparison between the sampled $\tilde{\tau}_{\rm est}$ and the actual $\tilde{\tau}_{\rm norm}$. 
Despite the many approximations under the estimator Eq.\;\ref{eq:tau_est}, it is capable of providing an order-of-magnitude estimate of $\tilde{\tau}_{\rm norm}$. 
Note that the estimator works even for $\tilde{\tau}_{\rm norm} = 1$ where the eikonal approximation fails and the whole `three-image zone' composes only one wide interference fringe. 

That the estimator Eq.\;\ref{eq:tau_est} is systematically biased low is mainly because we have neglected the second term on the right-hand side of Eq.\;\ref{eq:fringe}. This term has an opposite sign to the first term, and thus neglecting it leads to an under-estimation of $\tilde{\tau}_{\rm norm}$. 
When a dynamic spectrum of the three-image zone is available, one can derive the combination in the second term $\tilde{\tau}_{\rm norm} \nu (\tilde{\theta} - \tilde{\beta})$ from the secondary spectrum, and correct for this systematic bias (see Appendix.\;\ref{app:sec}).

Translating Eq.\;\ref{eq:tau_est} to observable quantities and taking $\tilde{\tau}_{\rm est}$ as $\tilde{\tau}_{\rm norm}$, we have
\eq{
\frac{V_{\rm sr}^2 }{ f_{\rm d} (1 - f_{\rm d}) D}   =  \frac{\it{c}}{\nu t \Delta t} \,,
}
i.e. observationally, from the interference pattern we can constrain a combination of $V_{\rm sr}$, $f_{\rm d}$ and $D$.
This is exactly the combination that enters Eq.\;\ref{eq:Ne_est}. 

Note that we have neglected the possible contamination from images other than the ones produced by the transiting lens. 
Actually, there could be a large number of additional images, considering the large number of arclets seen on some secondary spectra.
These images are in general much dimmer than the ones we consider here, and thus should not significantly affect our estimations in this paper that are based on bright features e.g. cusp, cusp characteristic, and caustic. However, the existence of additional images can make it hard to infer the detailed shape of the plasma lens using methods based on slight flux variations, and complicate the interference fringe pattern. Overall speaking, this makes it harder to derive the full shape of section A i.e. the central part of the lens when additional images exist since the information there is carried by de-magnified images.

\section{Conclusion}
\label{sec:conclusion}
Scintillation of compact radio sources has revealed intriguing facts about the ionized interstellar medium: 
the likely existence of discrete, highly anisotropic fine structures that act as plasma lenses, whose detailed properties and astrophysical correspondence are highly debatable.

We demonstrate that wide-band observations of dynamic spectra during plasma lensing events are valuable probes of the properties of these plasma lenses, complimentary to the pulsar secondary spectra
as well as the single-band light curves. 
To demonstrate this, we have performed accurate computation of dynamic spectra during plasma lensing events 
by direct evaluation of the Kirchhoff-Fresnel integral with the Picard-Liftschez method, and have compared them to the computations in the eikonal limit (i.e. stationary phase approximation). 
Dynamic spectra have been computed for well-motivated single-variable lenses at various distance normalizations.  

We show that the dynamic spectra during a plasma lensing event can constrain the plasma lens in several ways: 
The location of the cusp point constrains the typical size and strength of the lens; 
the shapes of the features e.g. spectral caustics on the dynamic spectra constrain the shape of the lens;
and the width of the interference pattern constrains a combination of the distances and the effective velocity.

Future wide-band observations of pulsars are expected to capture plasma microlensing events at sufficient dynamic range to enable such studies,
and shed more light on the mysterious fine-scale structure of the ionized interstellar medium. 

\section*{Data Availability Statements}
\noindent No new data were generated or analyzed in support of this research.

\section*{Acknowledgements}
\noindent We thank the organizers and speakers of the ``Gravity meets Plasma'' workshop held in Yunnan University in summer 2019 for introducing this field to us.
The \textsc{Python} codes we use in computing the Kirchhoff-Fresnel integral for 2D single-variable frequency-dependent lenses are based on Job Feldbrugge's Picard-Lefschetz integrator for 1D lenses implemented with \textsc{C++} which is publicly available. 
We are grateful to Xinzhong Er for reading the manuscript and for the helpful comments and suggestions, and to our referee Mark Walker for the top-quality referee report. 

\bibliographystyle{mnras}
\bibliography{bibliography}

\appendix

\section{Plasma microlensing light curve}
\label{app:light_curve}
\begin{figure}
    \includegraphics[width=0.48 \textwidth]{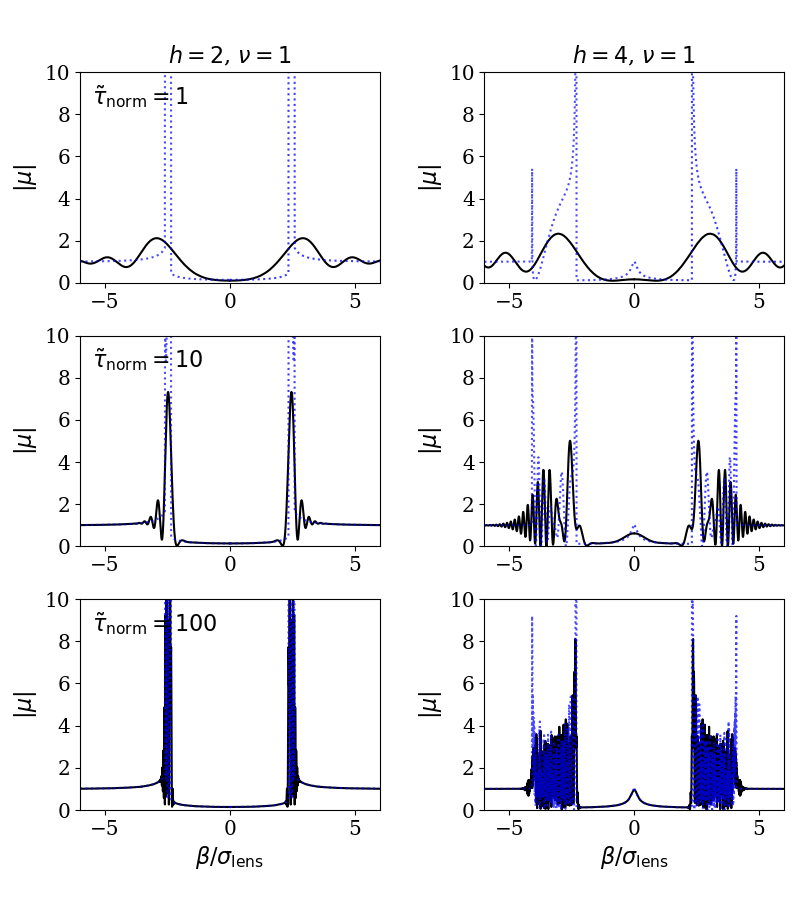} 
    \caption{Light curves during plasma microlensing by the $h=2$ (left panels) and $h=4$ (right panels) lenses at $\nu = 1$ GHz. 
    Distance normalization ($\tilde{\tau}_{\rm norm} = 1$, 10 and 100 GHz$^{-1}$ for the upper, middle, and lower panels) is essential in setting the shape of the light curve and the accuracy of the eikonal approximation (blue dashed lines) compared with the direction integration results of the Kirchhoff-Fresnel integral (black solid lines). 
    }
    \label{fig:lightcurve}
\end{figure}

\begin{figure}
    \includegraphics[width=0.48 \textwidth]{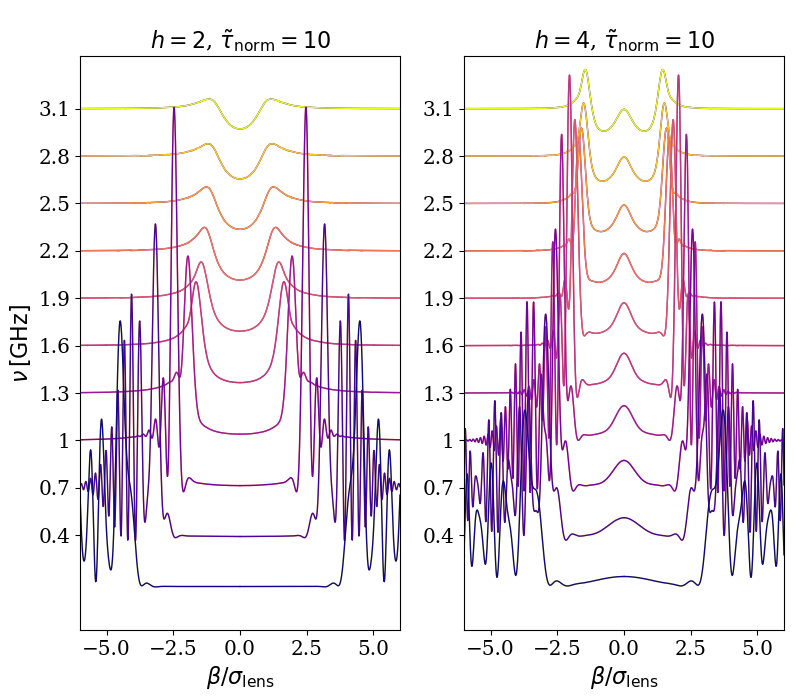} 
    \caption{Variation of plasma microlensing light curves shapes with observing frequency. The $h=2$ (left panels) and $h=4$ (right panels) lenses with a particular distance normalization $\tilde{\tau}_{\rm norm} = 10$ GHz$^{-1}$ are considered. }
    \label{fig:alllightcurve}
\end{figure}

In this appendix, we present plasma microlensing light curves extracted from the modeled dynamic spectra.

Fig.\;\ref{fig:lightcurve} presents light curves observed at a fixed frequency $\nu = 1$. 
As the distance normalization $\tilde{\tau}_{\rm norm}$ increases (from top panels to bottom ones), the light curve (black line) exhibits sharper caustics
and are better approximated by the eikonal limit computations (blue line).
The $h=4$ lens (right panels) features a local maxima in the light curves at source-lens alignment ($\beta=0$) in comparison to a total depletion for the $h=2$ (left panels) lens. 

Fig.\;\ref{fig:alllightcurve} demonstrates the frequency-dependency of the light curves. 
Light curves at lower frequencies have wider depletion zones since they are refracted more.
The U-shaped (W-shaped) light curves that are trademarks of the $h=2$ ($h=4$) lenses are evident at high frequencies.

\section{Location of the cusp point and the cusp characteristic}
\label{app:cusp}

\begin{figure}
    \includegraphics[width=0.4 \textwidth]{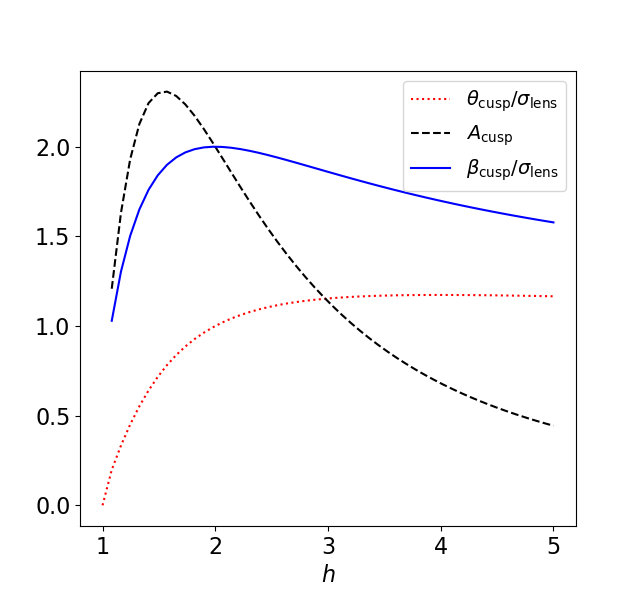}
    \centering
    \caption{The image position ${\theta_{\rm cusp}}/{\sigma_{\rm lens}}$ (red dotted line),  source position ${\beta_{\rm cusp}}/{\sigma_{\rm lens}}$ (blue solid line),  the amplitude of the cusp point $A_{\rm cusp}$ (black dashed line) as functions of model paramter $h$.}
    \label{fig:cusp}
\end{figure}

\begin{figure}
    \includegraphics[width=0.4 \textwidth]{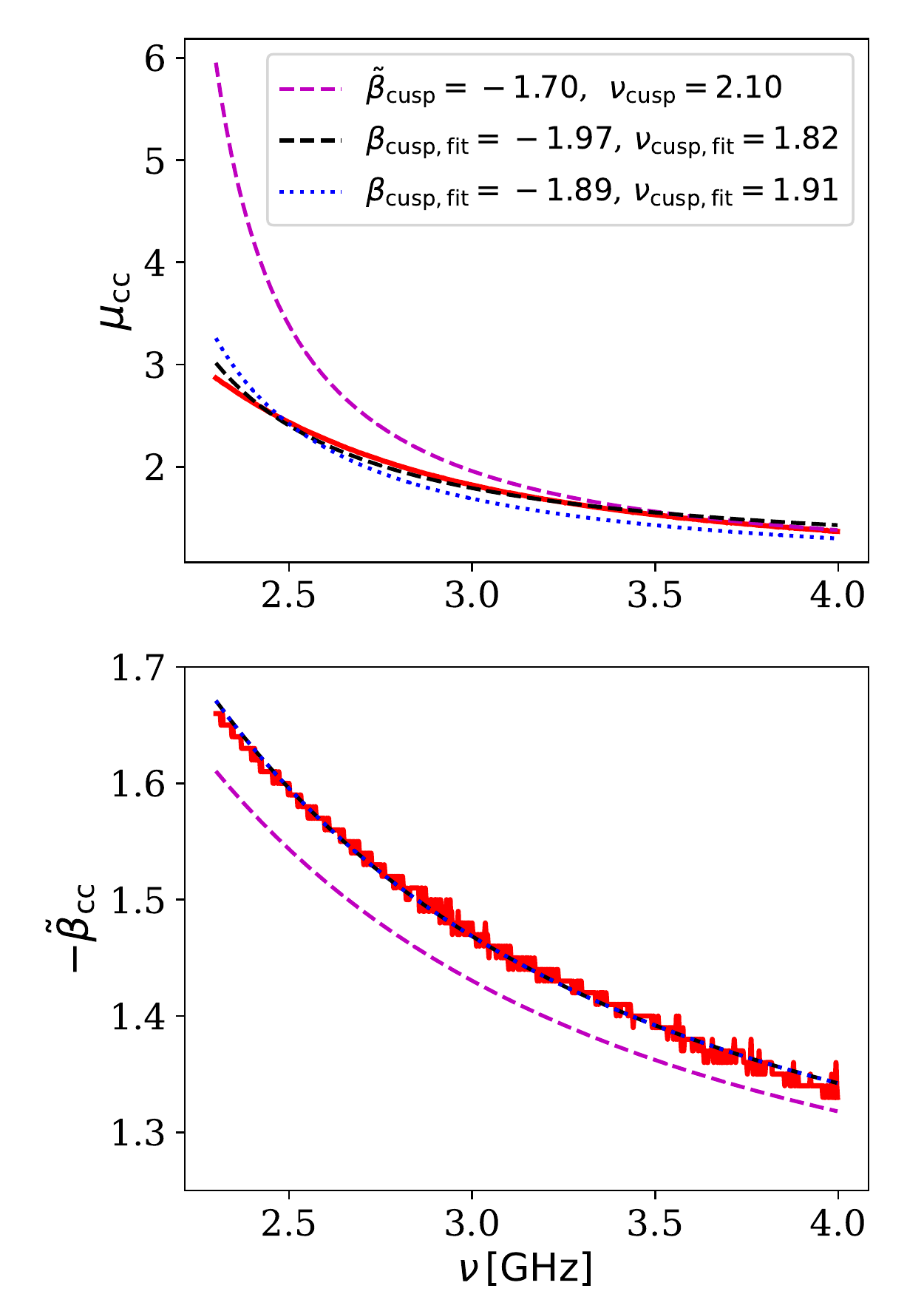}
    \centering
    \caption{Estimating cusp location $\tilde{\beta}_{\rm cusp}$ and $\nu_{\rm cusp}$ using measured location of the cusp characteristic $\tilde{\beta}_{\rm cc}$ and the magnification $\mu_{\rm cc}$ along it as functions of $\nu_{\rm cc}$(red solid lines) extracted from Fig.\;\ref{fig:caustics_est}. Magenta dashed lines are estimations based on Eqs.\;\ref{eq:beta_cc} and \ref{eq:mu_cc} using true parameter values in the eikonal limit. The black dashed and blue dotted lines are those using fitted parameters. In one of the fittings (black dashed lines), we have assumed an unknown normalization of measured $\mu_{\rm cc}$ to represent an estimation using a dynamic spectrum without an absolute flux calibration. Note that the fitted parameters depend on the selected range of the cusp characteristic. Their values would approach the true values when data far away from the cusp point are used as the eikonal approximations (magenta dashed lines) approach the measurements (red solid lines).}
    \label{fig:cusp_characteristic}
\end{figure}

The deflection angle and its derivative for our plasma lens model Eq.\;\ref{eq:rational_lens} are
\eq{
\label{eq:alpha}
\frac{\alpha}{\sigma_{\rm lens}} = -\frac{A  h (\theta / \sigma_{\rm lens})^{h-1}}{\bb{1 + (\theta / \sigma_{\rm lens})^{h}}^2}   \,,
}
and
\eq{
\label{eq:alphadot}    
 \dot{\alpha}  = \frac{2 A h^2 (\theta / \sigma_{\rm lens})^{2h-2}}{\bb{1 + (\theta / \sigma_{\rm lens})^{h}}^3} - \frac{A h (h-1) (\theta / \sigma_{\rm lens})^{h-2}}{\bb{1 + (\theta / \sigma_{\rm lens})^{h}}^2} \,.
}

The cusp point in the dynamic spectrum correspond to the maximum of $\dot{\alpha}$ as a function of $\theta$, 
which occurs at
\eqs{
\label{eq:theta_max}
\frac{\theta_{\rm cusp}}{\sigma_{\rm lens}} & = \bb{\frac{2 h^2 + \sqrt{3} \sqrt{h^4 - h^2} - 2}{h^2 + 3 h + 2}}^{1/h}    \\
 & \approx 1 \textrm{\ for\ }h \gtrsim 2  \,.
}
Inserting the exact expression into Eq.\;\ref{eq:alphadot}, we obtain
\eqs{
\dot{\alpha}_{\rm max} = & A (h+1)^2 (h+2) \left(h^2+3 h+2\right)^{2/h} \left(\sqrt{3} \sqrt{h^2-1}+h-1\right) \\
& \times \frac{\left(2 h^2+\sqrt{3} \sqrt{h^2-1} h-2\right)^{\frac{h-2}{h}}}{h \left(\sqrt{3} \sqrt{h^2-1}+3 h+3\right)^3} \,.
}
Specifically, $\dot{\alpha}_{\rm max} = 0.5 A$ for $h=2$ and$1.47 A$ for $h=4$.

The amplitude of the lens at the cusp point $A_{\rm cusp}$ should allow $\dot{\alpha} = 1$ (thus $\mu \to \infty$, cf. Eq.\;\ref{eq:constraint3}), and thus
\eqs{
A_{\rm cusp}  = & h (h+1) (h+2)^2 \left(h^2+3 h+2\right)^{-\frac{2}{h}-3} \\
& \times \frac{\left(\sqrt{3} \sqrt{h^2-1}+3 h+3\right)^3 \left(2 h^2+\sqrt{3} \sqrt{h^2-1} h-2\right)^{\frac{2}{h}-1}}{\sqrt{3} \sqrt{h^2-1}+h-1} .
}

Inserting this and Eq.\;\ref{eq:theta_max} into Eq.\;\ref{eq:alpha}, we obtain
\eq{
 \frac{\alpha_{\rm cusp}}{\sigma_{\rm lens}} =-\frac{\left(\sqrt{3} \sqrt{h^2-1}+3 h+3\right) \left(\frac{h^2+3 h+2}{2 h^2+\sqrt{3} \sqrt{h^2-1} h-2}\right)^{-1/h}}{(h+1) \left(\sqrt{3} \sqrt{h^2-1}+h-1\right)} \,,
}
and subquently,
\eqs{
\frac{\beta_{\rm cusp}}{\sigma_{\rm lens}} & = \frac{\theta_{\rm cusp} - \alpha_{\rm cusp}}{\sigma_{\rm lens}} \\
&  = \frac{(h+2) \left(\sqrt{3} \sqrt{h^2-1}+h+1\right) \left(\frac{2 h^2+\sqrt{3} \sqrt{h^2-1} h-2}{h^2+3 h+2}\right)^{1/h}}{(h+1) \left(\sqrt{3} \sqrt{h^2-1}+h-1\right)} \,.
}
See Fig.\;\ref{fig:cusp} for the magnitudes of $\alpha_{\rm cusp}$, $\beta_{\rm cusp}$, and $A_{\rm cusp}$ as functions of $h$. We have assumed $h \geq 1$ in the computations above.

These results imply that $\beta_{\rm cusp} \gtrsim \sigma_{\rm lens}$, i.e. the width of the depletion zone in terms of angular size directly reflects the typical size of the plasma lens $\sigma_{\rm lens}$. 
The amplitude of the lens at the cusp point $A_{\rm cusp}$ depends more sensitively on the lens shape, which is characterized by the outer slope $h$ in our model, but can be taken as $A_{\rm cusp} \sim 1$ for moderate $h$ values $1 \lesssim h \lesssim 5$ (See Fig.\;\ref{fig:cusp}). 
At $h \gg 1$, $|\beta_{\rm cusp}| / \sigma_{\rm lens}$ approaches unity, but $A_{\rm cusp}$ decreases in proportional to $h^{-2}$ since the (de-)focusing power of the lens depends on the second derivative of the lensing potential. Thus, measuring $\beta_{\rm cusp}$ and $\nu_{\rm cusp}$ allows for an estimation of the amplitude and the size of the plasma lens (cf. Eq.\;\ref{eq:Ne_est}).

The cusp characteristic is the one that passes through the cusp point and extends to the $\nu > \nu_{\rm cusp}$ region as the bright ridge on the dynamic spectrum. It shares the same image position $\theta_{\rm cusp}$ with the cusp point, and is described by 
\eq{
    \label{eq:beta_cc}
    \beta_{\rm cc} = \theta_{\rm cusp} - \frac{\nu_{\rm cusp}^2}{\nu^2} \alpha_{\rm cusp} \,.
}
The magnification along the cusp characteristic in the eikonal limit is 
\eq{
    \label{eq:mu_cc}
    \mu_{\rm cc} = \frac{1}{1 - \nu_{\rm cusp}^2 / \nu^2} \,.
}
Thus, measuring the cusp characteristic location $\beta_{\rm cc}(\nu)$ and magnification $\nu_{\rm cc}(\nu)$ enables the determination of $\theta_{\rm cusp}$, $\alpha_{\rm cusp}$ and $\nu_{\rm cusp}$, and consequently the amplitude and the size of the plasma lens. The estimated values deviate from the true values due to the limited validity of the eikonal approximation (Fig.\;\ref{fig:cusp_characteristic}), but would still suffice for an order-of-magnitude estimate of the plasma lens amplitude and size.

\section{Secondary spectrum in the three-image zone}
\label{app:sec}
\begin{figure}
    \centering
    \includegraphics[width=0.42 \textwidth]{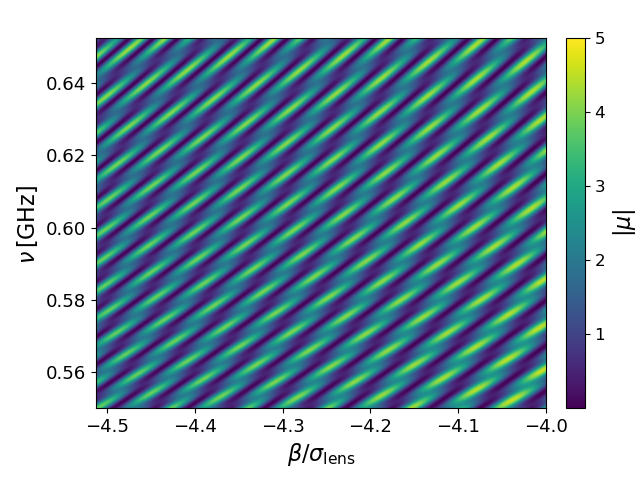}\\
    \includegraphics[width=0.42 \textwidth]{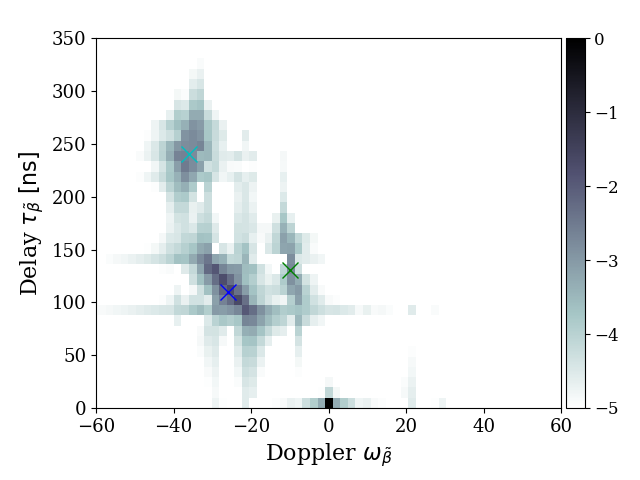}
    \caption{A fraction of the dynamic spectrum in the three-image zone of the lower left panel of Fig.\;\ref{fig:dyn} and its secondary spectrum showing the contribution of the three images. The blue and cyan crosses on the secondary spectrum shows the relative delay and doppler of the two fainter images with respect to the main image computed using the lensing framework at the central frequency and $\beta$ of the dynamic spectrum. The green cross shows those of the two fainter images with respect to each other.}
    \label{fig:fringe2sec}
\end{figure}  

In this paper, we have focused on information about the plasma lens directly from the dynamic spectrum. 
Yet, the common approach that has been proven successful is to derive information in the Fourier domain, i.e. from the secondary spectrum. This approach has its advantage since the power distribution on a secondary spectrum can be directly mapped to the doppler and delay of the images, and the data it requires is only a relatively small number of interference fringes.

Here, we examine the information contained in the secondary spectrum of a small section of the dynamic spectrum in the three-image zone (Fig.\;\ref{fig:fringe2sec}). The patches of concentrated power in the secondary spectrum arise from the interference among the images. One of the three images is the main image with $\mu \approx 1$ which we call image $a$. It is located at a position close to the source position $\theta_a \approx \beta$. The other two images $b$ and $c$ are de-magnified when the dynamic spectrum section is taken away from the caustics. Their interference with image $a$ leads to the two brighter patches in the secondary spectrum (the blue and cyan crosses in Fig.\;\ref{fig:fringe2sec}), and the interference between image $b$ and $c$ leads to the fainter patch around the green cross. 

The doppler $\omega_{\tilde{\beta}}$ and delay $\tau_{\tilde{\beta}}$ of an image $i = a, b , c$ are  
\eq{
    \omega_{\tilde{\beta}, i} \equiv \frac{1}{2\uppi}\frac{\partial \Phi}{\partial \tilde{\beta}} = - \frac{\tilde{\tau}_{\rm norm} \nu}{2\uppi} \br{\tilde{\theta}_i - \tilde{\beta}} \,,
}
\eq{
    \tau_{\tilde{\beta}, i} \equiv \frac{1}{2\uppi}\frac{\partial \Phi}{\partial \tilde{\nu}} =  \frac{\tilde{\tau}_{\rm norm}}{2\uppi} \bb{\frac{|\tilde{\theta}_i - \tilde{\beta}|^2}{2} + \psi(\tilde{\theta}_i)} \,,
}
Note that it is $+ \psi$ in the expression of the delay $\tau_{\tilde{\beta},i}$ in constrast to the $-\psi$ in that of the phase delay (Eq.\;\ref{eq:phase_delay}). This results from the $\nu^{-2}$ frequency dependence of the lensing potential $\psi$.

The doppler and delay of the patches on the secondary spectrum are the differences between those of the images. For example, the location of the blue cross in Fig.\;\ref{fig:fringe2sec} can be estimated as 
\eq{
    \omega_{\tilde{\beta}} = - \frac{\tilde{\tau}_{\rm norm} \nu}{2\uppi} \br{\tilde{\theta}_b - \tilde{\theta}_a} \,,
}
\eq{
    \tau_{\tilde{\beta}} =  \frac{\tilde{\tau}_{\rm norm}}{2\uppi} \bb{\frac{|\tilde{\theta}_b - \tilde{\beta}|^2 - |\tilde{\theta}_a - \tilde{\beta}|^2}{2} + \psi(\tilde{\theta}_b) - \psi(\tilde{\theta}_a)} \,.
}
Taking the approxmation $\theta_a \approx \beta$ and assuming $\tilde{\tau}_{\rm norm}$ is known, measuring the location of the patches on a secondary spectrum can yield an estimation of $\tilde{\alpha}_b = \tilde{\theta}_b - \tilde{\beta}$ and the lensing potential $\psi(\tilde{\theta}_b)$ relative to $\psi(\tilde{\beta})$. 

In the example we show here (Fig.\;\ref{fig:fringe2sec}), the contribution of the deflection potential to the delay i.e. the dispersive delay is comparable to the geometrical delay, and thus we do not expect the power distribution on the secondary spectrum to lie on a parabolic arc. This is because we are examining an interference pattern produced near transit i.e. the separation between the lens and the source is only a few times the width of the lens. In another word, we are considering a situation where the arclets are very close to the origin of a secondary spectrum. Observations of arclets so far have suggested the dominance of the geometrical delay, which means that the images corresponding to the observed arclets are far from the source compared to the size of the lens. This also implies that the observing frequency in those observations is not very close to the cusp frequency, since at $\nu$ around $\nu_{\rm cusp}$ that we consider in this paper, the three-image zone does not extend to a very large source-lens separation $\beta$. 

The doppler and delay of the images depend on the observing frequency and the source location. In a dynamic spectrum with a not-too-small bandwidth and time duration, the change of image doppler and delay across the dynamic spectrum leads to a smearing of the power distribution that is evident in Fig.\;\ref{fig:fringe2sec}. In principle, one can measure this change by e.g. constructing secondary spectra for fractions of the original dynamic spectrum at different central frequencies and/or times. This would allow one to sample the deflection angle $\tilde{\alpha}$ and the lensing potential $\psi$ at different image locations $\tilde{\theta}$. Note that these images are de-magnified i.e. with $\dot{\alpha} < 0$, they lie on section A of the $\alpha - \dot{\alpha}$ diagram. In fact, measuring the inference pattern of the de-magnified images with the bright main image close to transit is perhaps the best way to constrain section A of the $\alpha - \dot{\alpha}$ diagram and thus the central part of a diverging plasma lens. 


\end{document}